# Propensity Straight-Through Gradients for Discrete Stochastic Systems

Jose M. G. Vilar[1,2,*] and Leonor Saiz[3,*]
[1]Biofisika Institute (CSIC, UPV/EHU), University of the Basque Country, P.O. Box 644, 48080 Bilbao, Spain
[2]IKERBASQUE, Basque Foundation for Science, 48011 Bilbao, Spain
[3]Department of Biomedical Engineering, University of California, 451 East Health Sciences Drive, Davis, CA 95616, USA

[*] To whom correspondence should be addressed: j.vilar@ikerbasque.org or lsaiz@ucdavis.edu

## Abstract

We introduce the propensity straight-through (PST) estimator for gradient-based learning through exact simulations of continuous-time Markov chains (CTMCs). These models describe discrete stochastic dynamics across physics, chemistry, and biology, but their categorical event choices prevent ordinary differentiation. PST preserves exact forward trajectories and differentiates, in the backward pass, the normalized reaction propensities the simulator already computes. The affine state update makes this rule exact for the one-step conditional-mean sensitivity. We derive the per-step discrepancy for one-step sensitivities composed along nonlinear trajectories in closed form, show that it vanishes for affine downstream dependence, and bound it quadratically in the stoichiometric jump size. Across reversible dimerization, a genetic oscillator, a 50-task repressilator suite, and experimental patch-clamp recordings, PST recovers parameters to within 0.06–1.7% and converges 2.4–3.0-fold faster than the Gumbel-Softmax straight-through estimator at matched computational cost. At deep-learning scale, PST trains all 203,796 parameters of a stochastic reaction network using exact discrete events from initialization. By differentiating an exact conditional mean, PST eliminates reaction-selection temperature tuning and provides a scalable route to learning through exact stochastic dynamics.

## Introduction

Discrete stochasticity is inherent to physical, chemical, and biological systems whenever individual events remain significant at the scale of the observed behavior. Molecular reactions, conformational transitions, ion-channel openings, and population changes generate fluctuations that determine switching, extinction, oscillatory variability, and rare transitions rather than merely perturb an underlying deterministic trajectory (1–12). Continuous-time Markov chains (CTMCs) provide a natural framework for representing such systems, with event rates specifying both the stochastic timing and identity of discrete state changes. Exact event-driven simulation is therefore particularly important in regimes where diffusion or deterministic approximations can mask the consequences of individual transitions.

Inference and design increasingly require these mechanistic models to be embedded within optimization procedures. Experimental data may constrain many kinetic parameters simultaneously, while inverse-design and machine-learning applications reach parameter spaces far beyond those accessible to parameter-by-parameter search. Gradient-based optimization is the natural route to such problems, and automatic differentiation has made it routine for deterministic models at scales of millions of parameters.

Exact discrete-event simulation, epitomized by the Gillespie stochastic simulation algorithm (SSA) and the Bortz–Kalos–Lebowitz (BKL) algorithm, generates exact sample paths of continuous-time Markov jump processes by alternating exponential waiting-time sampling with categorical reaction selection (13, 14). The approach is used in statistical physics, stochastic chemical kinetics, systems biology, queueing, epidemiology, and other applications governed by competing event rates. Gradient-based optimization through such a trajectory is challenging because the selected event is discrete and blocks ordinary pathwise differentiation.

The waiting time is straightforwardly differentiable. Conditional on the current state $\mathbf{x}$, it can be written as $\Delta t = -\log u / a_0(\mathbf{x}; \boldsymbol{\theta})$, where $u$ is uniform and $a_0$ is the total propensity. The reaction choice, however, is a categorical sample, a piecewise constant function of its probabilities, and its derivative is consequently zero almost everywhere. The piecewise-constant dependence of the categorical sample on its probabilities does not preclude the applicability of score-function, or likelihood-ratio, estimators (15, 16), which differentiate the probability of the sampled reaction rather than the reaction realization itself, but it prevents ordinary pathwise differentiation through the categorical realization.

We show that the probabilities already used to select exact stochastic reactions directly supply a learning rule for mechanistic models. The affine structure of the SSA state update provides an exact conditional-mean foundation for this construction. If $\mathbf{e}_J$ is the one-hot indicator of the selected reaction, equivalently the corresponding reaction-selection vector, and $\mathbf{S}$ contains the stoichiometric change vectors as columns, the state update is affine, $\mathbf{x}^+ = \mathbf{x} + \mathbf{S}\mathbf{e}_J$ . The conditional mean of $\mathbf{e}_J$ is the normalized propensity vector $\boldsymbol{\pi}$. Therefore, differentiating $\boldsymbol{\pi}$ gives the exact derivative of the one-step conditional mean state. We use this derivative in the backward pass of automatic differentiation while leaving the waiting-time sampling and hard reaction selection unchanged, defining the propensity straight-through (PST) estimator. The affine structure is essential because the exactness applies to the one-step conditional mean and affine post-reaction observables.

Differentiable Gillespie approaches have also been developed by smoothing discontinuous operations in the forward dynamics (17). Exact-forward differentiable SSA implementations instead retain hard categorical reaction selection and use the Gumbel-Softmax straight-through estimator (GS-ST), differentiating a temperature-dependent relaxed realization only in the backward pass (18, 19). The relaxation temperature is a backward-pass hyperparameter rather than a property of the stochastic model, and the GS-ST settings represented across existing applications and the comparisons considered here vary markedly between systems. PST removes both this temperature selection and the Gumbel perturbation from the backward reaction-selection rule. For PST, the normalized propensities already calculated by the simulator supply the conditional-mean sensitivity propagated through the exact reaction trajectory.

The practical challenge at high parameter dimension is to obtain useful gradients for all parameters at a feasible simulation cost. Standard parameterwise coupled finite-difference and Poisson-path constructions incur additional simulation or sensitivity-calculation costs for each parameter, as discussed in Ref. (18). Score-function estimators can compute gradients for all parameters jointly, but their variance can require large trajectory ensembles for useful optimization updates (15, 16). These are distinct computational and sampling obstacles. The scalable GS-ST framework addresses this learning problem by combining massively parallel SSA with reverse-mode differentiation through a straight-through surrogate (18). PST replaces the relaxed reaction-selection derivative with an exact conditional-mean sensitivity, combining this computational scalability with a temperature-free backward rule and an explicit trajectory-sensitivity analysis.

PST enables exact-event training from initialization at deep-learning parameter scale. Our previous work (18) established massively parallel GS-ST optimization and demonstrated operation at 203,796 parameters. That classifier used soft-forward sampling for the first 16 of 80 epochs before switching to hard-forward GS-ST. Here the same 203,796-parameter architecture is trained with exact categorical sampling from the first epoch, without a reaction-selection temperature or a soft-forward warm-up, thus providing the first demonstration of a neural network realized as a stochastic chemical reaction system trained throughout by a gradient rule on exact discrete events alone.

The mathematical starting point is the expected next state conditional on the current state, whose sensitivity follows exactly from the affine SSA update. PST supplies a trajectory learning rule by propagating this sensitivity through state-dependent reactions, waiting times, and observable readouts. At the categorical level, this construction is closely related to marginal-based straight-through estimation (20), which retains discrete forward samples and differentiates exact sampling marginals. PST composes this local rule along a trajectory of chained categorical events: each reaction updates the state, the state sets the propensities of the next draw, and differentiable waiting times intervene between draws. Therefore, the central contribution here is the resulting construction and analysis for learning through exact stochastic reaction trajectories, evaluated from mechanistic inference to high-dimensional training. PST belongs to the broader family of straight-through estimators (21) and complements established sensitivity-analysis approaches for discrete stochastic systems, including likelihood-ratio, finite-difference and coupling, and pathwise approaches, with unbiased constructions among them (15, 16, 22–26).

We establish PST as a direct, scalable learning rule for exact discrete stochastic dynamics, with analytical foundations and demonstrated gains in inference performance. We demonstrate exactness for the one-step conditional mean state and affine post-reaction observables, identify the local reaction-selection discrepancy as a finite-difference-to-directional-derivative replacement propagated across events, derive a local bound involving curvature, jump sizes, and probability sensitivities, and connect its small-jump structure to classical expansions of Markov jump processes (27–30). For weighted least-squares population mean-matching objectives, we further show that the PST update field vanishes wherever the simulated population mean equals the target. We then evaluate PST in reversible dimerization, a nonlinear genetic oscillator (31), a two-channel gating model fitted to experimental recordings (32, 33), and a released 50-task repressilator suite (34), and apply it to the established 203,796-parameter stochastic gene-regulatory classifier (18). For the ion-channel model, we further test optimization robustness across initial values of the kinetic parameters and stochastic seed blocks under a substantially reduced training ensemble.

# Results

## PST construction and exact one-step sensitivity

PST preserves exact SSA sampling in the forward pass while replacing the derivative through categorical reaction selection by the derivative of the normalized propensities. Throughout, *exact-forward* refers to this unmodified SSA sample path, and *one-step exact* refers specifically to the conditional-mean sensitivity derived below.

Consider a reaction network with state $\mathbf{x} \in \mathbb{Z}_{\geq 0}^{N}$, reaction channels $j = 1, \ldots, M$, propensities $a_j(\mathbf{x};\boldsymbol{\theta}) \geq 0$, and stoichiometric changes $\boldsymbol{\nu}_j \in \mathbb{Z}^N$. For a non-absorbing state, the total and normalized propensities are given by $a_0(\mathbf{x};\boldsymbol{\theta}) = \sum_{j=1}^{M} a_j(\mathbf{x};\boldsymbol{\theta})$ and $\pi_j(\mathbf{x};\boldsymbol{\theta}) = a_j(\mathbf{x};\boldsymbol{\theta})/a_0(\mathbf{x};\boldsymbol{\theta})$, respectively. With $\mathbf{S} = [\boldsymbol{\nu}_1 \ \cdots \ \boldsymbol{\nu}_M]$, direct SSA samples $J \sim \mathrm{Categorical}(\boldsymbol{\pi})$ and updates the state as $\mathbf{x}^{+} = \mathbf{x} + \mathbf{S}\mathbf{e}_J$. PST leaves this hard reaction selection and the standard exponential waiting-time sampling unchanged, modifying only the derivative through the reaction indicator:

$$\mathbf{y}_{\mathrm{PST}} = \boldsymbol{\pi} + \mathrm{stopgrad}(\mathbf{e}_J - \boldsymbol{\pi}).$$

Numerically, the forward state remains an exact discrete SSA state, $\mathbf{y}_{\text{PST}} = \mathbf{e}_J$, while automatic differentiation gives $\nabla_{\boldsymbol{\theta}} \mathbf{y}_{\text{PST}} = \nabla_{\boldsymbol{\theta}} \boldsymbol{\pi}$ (Figure 1). Full TensorFlow implementation details, including tensor conventions, categorical sampling, absorbing-state masks, and framework-specific code, are given in SI Appendix, SI Materials and Methods.

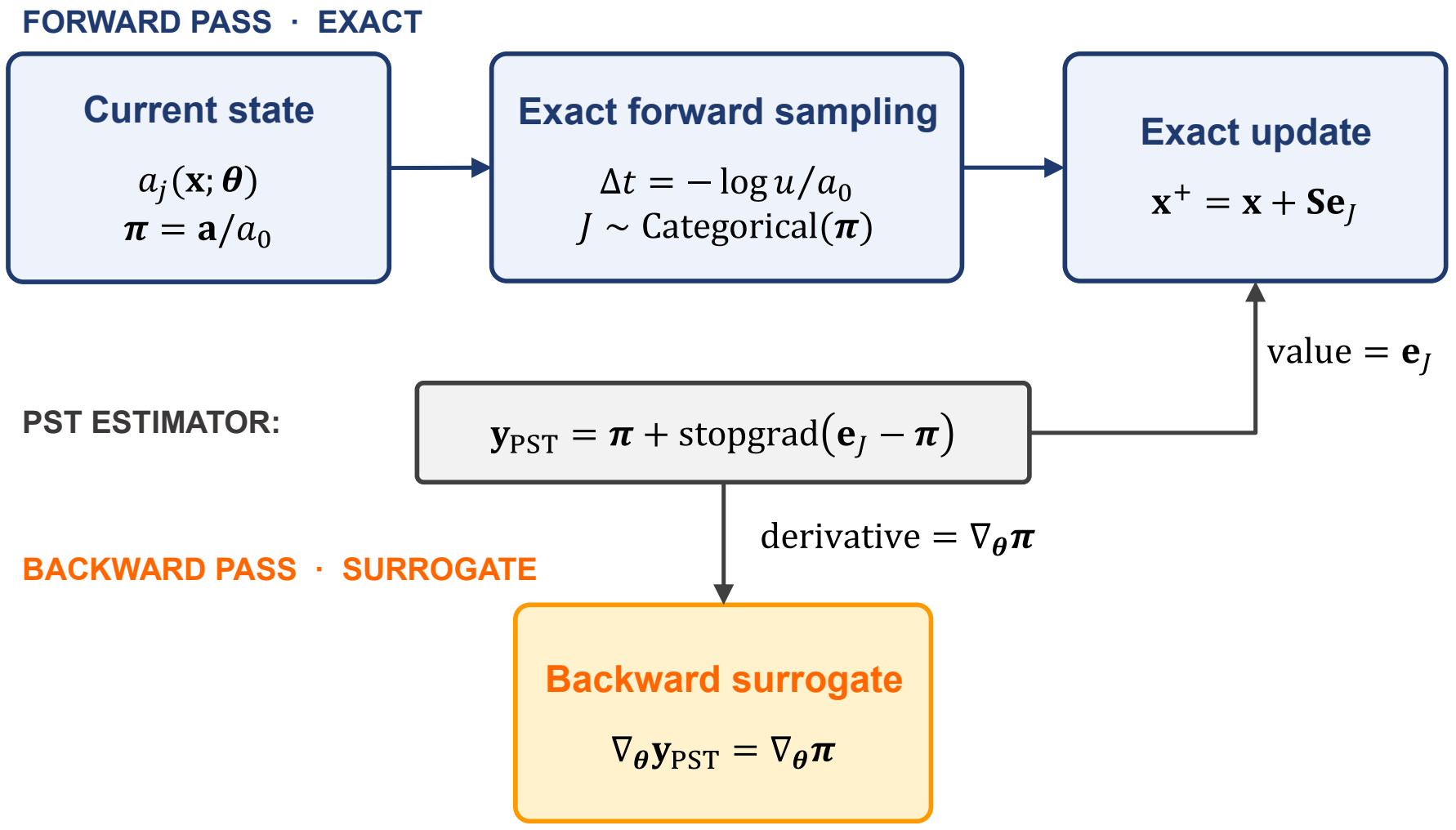

**Figure 1. The propensity straight-through (PST) estimator at a single SSA step.** The waiting time $\Delta t = -\log u / a_0$ with $u \sim \text{Uniform}(0,1)$ and the reaction index $J \sim \text{Categorical}(\boldsymbol{\pi})$ are sampled exactly from the normalized propensities $\boldsymbol{\pi} = \mathbf{a}/a_0$, $a_0 = \sum_j a_j$, and the hard update $\mathbf{x}^+ = \mathbf{x} + \mathbf{S}\mathbf{e}_J$ is retained, with $\mathbf{e}_J$ the one-hot indicator of the selected reaction and $\mathbf{S}$ the matrix of stoichiometric change vectors. PST replaces the reaction indicator by $\mathbf{y}_{\text{PST}} = \boldsymbol{\pi} + \text{stopgrad}(\mathbf{e}_J - \boldsymbol{\pi})$, whose value is $\mathbf{e}_{\text{J}}$ and whose derivative is $\nabla_{\boldsymbol{\theta}} \boldsymbol{\pi}$. The forward trajectory is therefore unchanged, while the backward pass returns the exact one-step conditional-mean sensitivity $\nabla_{\boldsymbol{\theta}} \mathbb{E}[\mathbf{x}^+ \mid \mathbf{x}] = \mathbf{S} \nabla_{\boldsymbol{\theta}} \boldsymbol{\pi}$.

For a fixed nonabsorbing state $\mathbf{x}$, $\mathbb{E}[\mathbf{e}_J \mid \mathbf{x}] = \boldsymbol{\pi}$. The affine update therefore gives

$$\mathbb{E}[\mathbf{x}^+ \mid \mathbf{x}] = \mathbf{x} + \mathbf{S}\boldsymbol{\pi}, \qquad \nabla_{\boldsymbol{\theta}} \mathbb{E}[\mathbf{x}^+ \mid \mathbf{x}] = \mathbf{S} \nabla_{\boldsymbol{\theta}} \boldsymbol{\pi}.$$

The PST backward pass returns exactly this derivative for the one-step mean and for, more generally, every affine post-reaction observable $\varphi(\mathbf{z}) = \mathbf{c}^{\mathsf{T}} \mathbf{z} + d$, for which $\nabla_{\boldsymbol{\theta}} \mathbb{E}[\varphi(\mathbf{x}^+) \mid \mathbf{x}] = \mathbf{c}^{\mathsf{T}} \mathbf{S} \nabla_{\boldsymbol{\theta}} \boldsymbol{\pi}$. Consequently, for a one-step squared-error objective defined on the population mean, PST gives the exact parameter gradient by the chain rule.

Conditional on the fixed current state, this one-step affine derivative does not depend on the realized reaction index and has zero reaction-selection variance. Complete-trajectory variance remains because the gradient also depends on earlier states, later reactions, waiting times, and the objective.

## Local approximation and propagation across events

For a nonlinear downstream observable, PST replaces finite differences among reachable states with directional derivatives. This local approximation can arise at a single categorical node and is propagated when the rule is composed along a trajectory. For a post-reaction state $\mathbf{z}$, $V(\mathbf{z}; \boldsymbol{\theta})$ denotes the continuation value, the conditional expectation of the downstream loss or of a scalar observable used to construct it. The exact process evaluates $V$ only at reachable states $\mathbf{z} = \mathbf{x} + \boldsymbol{\nu}_j$. In the PST backward pass, gradients are propagated through the post-reaction state, requiring derivatives of $V$ with respect to $\mathbf{z}$. That derivative is supplied by the continuous extension defined by the differentiable computational graph used for backpropagation. Holding the current state $\mathbf{x}$ fixed, the expected downstream objective is $F(\boldsymbol{\theta}) = \sum_{j=1}^{M} \pi_j\,(\mathbf{x};\boldsymbol{\theta}) V\big(\mathbf{x}+\boldsymbol{\nu}_j;\boldsymbol{\theta}\big)$.

The exact contribution from changing the current reaction probabilities can be written, after subtracting the common baseline $V(\mathbf{x})$, as

$$\mathbf{g}_{\text{exact}} = \sum_{j=1}^{M} (\nabla_{\boldsymbol{\theta}}\pi_j)\,\big[V\big(\mathbf{x}+\boldsymbol{\nu}_j\big) - V(\mathbf{x})\big].$$

The exact contribution is therefore weighted by finite differences between reachable states. PST retains the hard post-reaction state but propagates $\mathbf{S}\nabla_{\boldsymbol{\theta}}\boldsymbol{\pi}$ through the downstream computational graph. Writing the probability-weighted post-reaction gradient as $\overline{\nabla_{\mathbf{x}}V}(\mathbf{x}) = \sum_{\ell=1}^{M} \pi_\ell\, \nabla_{\mathbf{x}}V(\mathbf{x}+\boldsymbol{\nu}_\ell)$, the expected PST contribution is

$$\mathbf{g}_{\text{PST}} = \sum_{j=1}^{M} \big[\boldsymbol{\nu}_j^{\top}\overline{\nabla_{\mathbf{x}}V}(\mathbf{x})\big]\,\nabla_{\boldsymbol{\theta}}\pi_j.$$

PST thus replaces each exact finite difference by a probability-weighted directional derivative. The local discrepancy is

$$\mathbf{g}_{\text{exact}} - \mathbf{g}_{\text{PST}} = \sum_{j=1}^{M} \big[V\big(\mathbf{x}+\boldsymbol{\nu}_j\big) - V(\mathbf{x}) - \boldsymbol{\nu}_j^{\top}\overline{\nabla_{\mathbf{x}}V}(\mathbf{x})\big]\,\nabla_{\boldsymbol{\theta}}\pi_j.$$

If $V$ is twice continuously differentiable, the local discrepancy is bounded in terms of downstream curvature, stoichiometric jump sizes, and reaction-probability sensitivities (SI Appendix, Eq. (S35)) and vanishes when $V$ is affine. The bound concerns the reaction-selection contribution at the current categorical node.

The finite difference in $\mathbf{g}_{\text{exact}}$ is the same jump increment that appears in the backward generator of a Markov jump process. PST therefore has a small-jump interpretation at the level of reaction-selection sensitivity: it leaves the forward jump process exact and replaces only this backward sensitivity contribution with a local directional rule (27–30). The SI Appendix shows that the leading local discrepancy is quadratic under the smooth small-jump expansion and relates the approximation to classical small-jump and system-size expansions (29, 30).

The discrepancy expression makes the local approximation explicit. The discrepancy vanishes identically when $V$ is affine on the relevant neighborhood; this is a sufficient condition, and nonlinearity need not produce a nonzero discrepancy. The curvature bound is inherently local because the relevant curvature depends on the downstream objective, the remaining horizon, and the continuous extension defined by the backward computation. Over longer trajectories, the same finite-difference-to-directional-derivative replacement occurs at each categorical node, and the local discrepancies may accumulate, cancel, change sign, or be damped by the dynamics. The full derivation, sufficient exactness conditions, worked two-step example, waiting-time extension, backward jump-operator and system-size interpretation, and mean-matching stationarity result are given in SI Appendix, Supplementary Theory.

The four inference benchmarks considered here span complementary regimes, covering parameter recovery in a simple reaction network, learning from segments of nonlinear stochastic dynamics, inference from strongly discrete experimental data, and recovery across an externally specified fixed 50-task suite. Detailed model definitions, inference protocols, and statistical procedures are provided in Materials and Methods; the sections below focus on the resulting comparisons. A fifth calculation tests the same reaction-selection rule on an established architecture at deep-learning parameter scale.

## Improved parameter recovery in reversible dimerization

PST improves reversible-dimerization parameter recovery under the protocol used in Ref. (18). For the representative condition $k_2 = 0.32$, exact trajectories generated from the PST-inferred parameters reproduce the target ensemble means of all three species (Figure 2A). Both inferred rates converge rapidly under the matched protocol (Figure 2B). Across the eight conditions, the recovered $k_2$ values from both estimators lie essentially on the identity line (Figure 2C). The condition-level two-parameter mean absolute percentage error (MAPE), averaged across the eight conditions, is $0.060\%$ for PST and $0.093\%$ for GS-ST (Figure 2D). This lower mean error is obtained without the reaction-selection temperature schedule used by GS-ST.

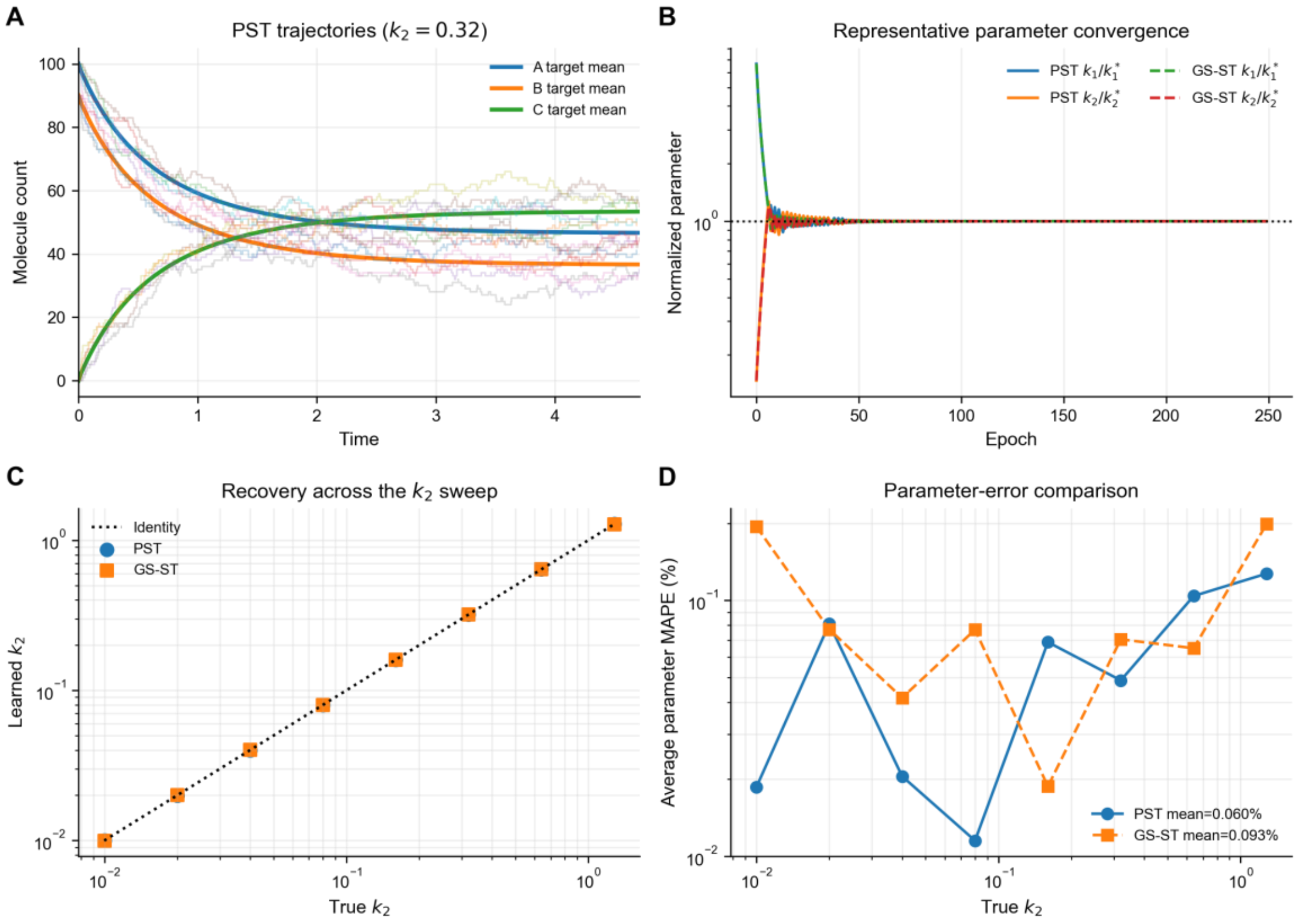

**Figure 2. Propensity straight-through differentiation in reversible dimerization. A**, Target ensemble means and representative exact trajectories generated with PST-inferred parameters for $k_2 = 0.32$. **B**, Representative parameter convergence for PST and GS-ST under the matched protocol. **C**, Recovered $k_2$ across the eight conditions; the dotted line denotes exact recovery. **D**, Condition-level two-parameter MAPE across the eight conditions.

## Faster and more accurate inference in a genetic oscillator

PST improves matched-horizon parameter accuracy and accelerates convergence in the nonlinear genetic oscillator (Figure 3B–D). Five kinetic parameters are inferred from short trajectory segments under the protocol used in Ref. (18). The ten GS-ST runs reported previously (18) are used as the comparison baseline rather than rerunning GS-ST. At 3000 epochs, the global pooled-vector MAPE, calculated by pooling terminal parameter samples across runs before forming one aggregate estimate, is 1.69% for PST and 1.23% for GS-ST, with the latter equal to the published value to the displayed precision. Exact trajectories generated with the PST aggregate parameters reproduce the characteristic amplitudes and waveform statistics of the activator, repressor, and activator–repressor complex (Figure 3A). Small parameter differences accumulate as phase shifts over successive cycles, making sustained phase locking unlikely between independently simulated stochastic oscillators over the full window.

At the matched 1000-epoch horizon, the median run-wise MAPE, calculated separately for each run before taking the median, is 1.70% for PST and 2.58% for GS-ST. At 3000 epochs, the corresponding values are 1.71% and 1.51% . The normalized parameter estimates have overlapping run-level distributions at both horizons (Figure 3D). The training histories show that PST remains nearly unchanged after epoch 1000, whereas GS-ST continues evolving (Figure 3B).

PST and GS-ST can be implemented through a common exact-forward code path that differs only in the backward reaction-selection rule. For cost matching, the same softmax implementation represents PST by removing the Gumbel perturbation and using unit softmax scaling, which returns the normalized propensities (SI Appendix, Section S8). This is an implementation identity rather than a GS-ST temperature limit: with the Gumbel perturbation retained, GS-ST differentiates a relaxed random sample at any temperature, whereas PST differentiates the categorical mean itself. In a GPU timing benchmark using matched hard Gumbel-Max sampling, the ratio of median PST to GS-ST optimization-step times was 1.002 (paired-bootstrap 95% interval 0.943–1.038; 18 repeats), supporting the cost-matched comparison (SI Appendix, Section S8). For sustained error below 3%, all ten runs in both arms meet the criterion. The median threshold epochs are 262.5 for PST and 782.5 for GS-ST, giving a GS-ST/PST ratio of median threshold epochs of approximately 3.0 (bootstrap 95% interval 2.4–3.5) (Figure 3C). Thus, under matched per-epoch computational cost, PST reaches the sustained 3% threshold in approximately one third of the optimization time required by GS-ST.

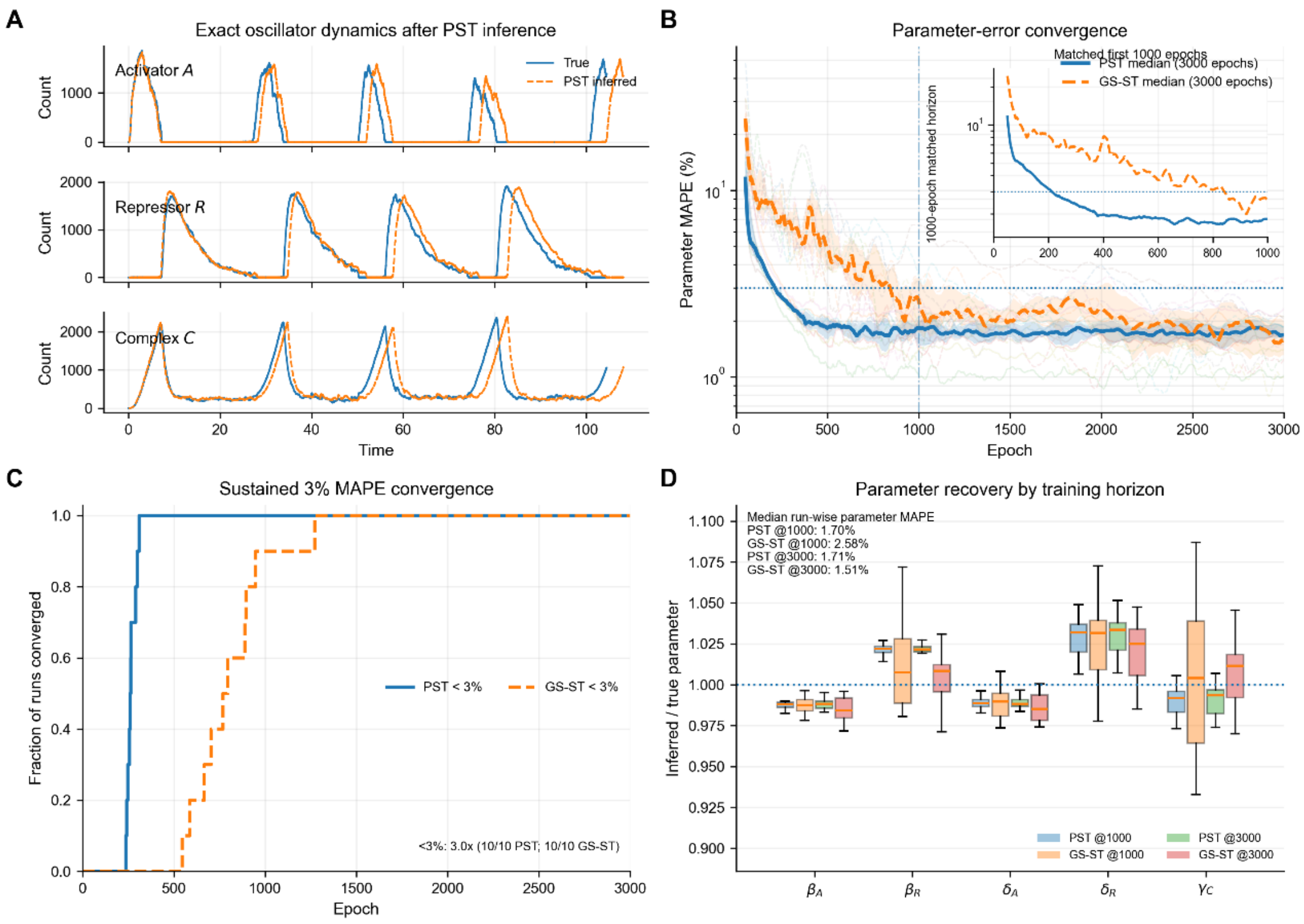

**Figure 3. Parameter inference in the stochastic genetic oscillator. A**, Exact trajectories generated with the true and PST-inferred parameters. **B**, Epoch-indexed parameter-MAPE histories over ten PST and ten GS-ST runs; thick curves are medians, faint curves are individual runs, shaded regions are interquartile ranges, and the vertical line marks the matched 1000-epoch horizon. These histories are distinct from the terminal-25% window summaries used for the horizon-specific estimates in panel **D** and the text. **C**, Fractions of runs reaching and sustaining the $3\%$ threshold under the trailing 50-epoch geometric-mean and 50-consecutive-epoch criterion. **D**, Run-wise parameter estimates normalized by their generating values at 1000 and 3000 epochs.

## Faster inference of experimental ion-channel kinetics

The ion-channel benchmark tests PST against experimental data in a strongly discrete regime, where a single opening or closing event changes the observable by half its full range. PST accurately reproduces the experimental mean trace (18). Exact-SSA validation produces visually overlapping mean traces, with $R^2 = 0.987$ and normalized root-mean-square error (NRMSE) $3.49\%$ for PST and $R^2 = 0.987$ and NRMSE $3.48\%$ for GS-ST (Figure 4A). Representative exact trajectories jump among zero, one, and two open channels (Figure 4B). The inactivated state is absorbing. The accurate fit demonstrates that PST is effective even when individual stoichiometric events are macroscopically large relative to the measured signal.

PST reaches its terminal-loss regime about twice as quickly as GS-ST (Figure 4C). In the single saved matched-protocol pair (18), a post hoc stabilization diagnostic gives epoch 75 for PST and epoch 160 for GS-ST, a GS-ST/PST stabilization-epoch ratio of 2.13. The stabilization-epoch ratio remains close to two across the tolerance bands reported in SI Appendix, Table S3.

PST also reproduces the pooled probabilities of observing zero, one, or two open channels (Figure 4D), providing an additional validation beyond agreement of the mean trace.

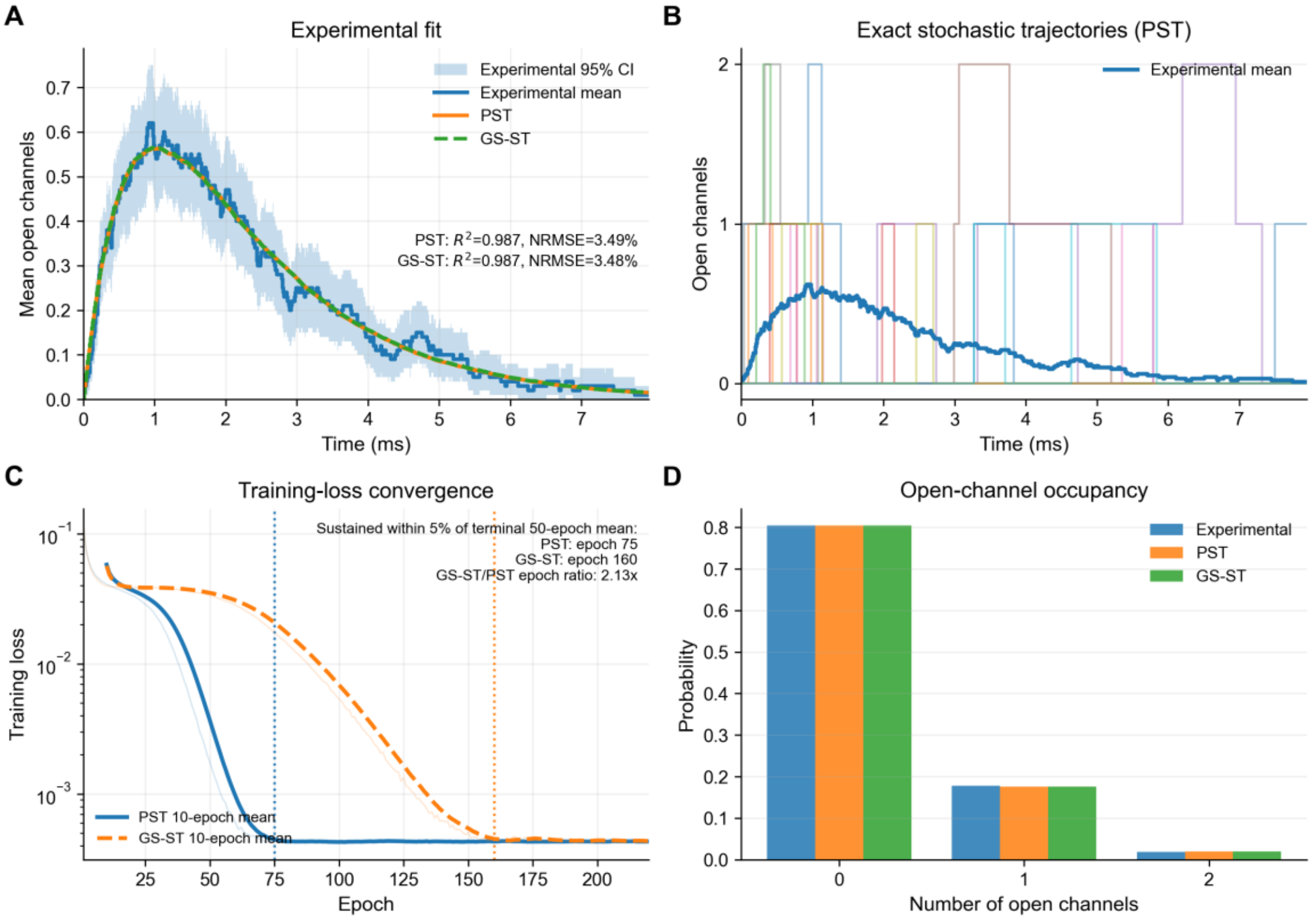

**Figure 4. Inference of ion-channel gating from patch-clamp recordings. A**, Experimental mean open-channel count with its pointwise 95% confidence interval and the means predicted by exact SSA trajectories using PST- and GS-ST-inferred rates. **B**, Representative exact PST trajectories, restricted to zero, one, or two open channels. **C**, Training losses for one PST and one GS-ST realization under the matched protocol. Faint curves show per-epoch losses and thick curves show trailing 10-epoch means. Vertical dotted lines mark the first epoch after which the moving mean remains within 5% of the final 50-epoch mean loss (epoch 75 for PST and epoch 160 for GS-ST); training continues for 400 epochs. **D**, Experimental and simulated open-channel occupancy distributions.

The convergence advantage persists across initial kinetic-parameter values and stochastic seed blocks with a more than 25-fold smaller training ensemble (Figure 5). The GS-ST temperature schedule and the matched learning-rate schedule were unchanged from the Figure 4 comparison. Ion-channel inference was repeated with 10,000 model trajectories per epoch, 40 SSA event slots per trajectory, and 400 epochs, using ten sets of initial kinetic-parameter values in each of three stochastic seed blocks (Figure 5).

To compare convergence on a common absolute scale, each training-loss history was smoothed with a trailing 10-epoch median and evaluated against a shared low-loss target defined as twice the pooled terminal median loss across both estimators. The convergence epoch was the first epoch at which the smoothed loss remained below this target for 10 consecutive epochs. PST reached the common target at a median of 36.0 epochs, compared with 90.5 epochs for GS-ST (Figure 5A,B), corresponding to a median paired GS-ST/PST epoch ratio of 2.41 (interquartile range, 2.18–2.69). PST reached the target earlier in all 30 matched comparisons across initial parameter values and seed blocks. The result was consistent across the three stochastic seed blocks, which gave median paired ratios of 2.42, 2.41, and 2.42, with PST earlier for all ten sets of initial kinetic-parameter values in each block.

The separation was robust to the convergence criterion. Across common targets from four times to 1.1 times the pooled terminal loss scale, the median paired GS-ST/PST epoch ratio remained between 2.36 and 3.16 (Figure 5C). All 30 matched comparisons reached the three least stringent targets. As the target approached the terminal loss scale, the number of comparisons in which both estimators crossed within 400 epochs decreased to 27, 22, and 17. Among the comparisons with finite crossing epochs for both estimators, PST was earlier at every tested target. At the primary target, changing the required persistence from 5 to 10 or 25 epochs left the median ratio unchanged at 2.41.

Independent exact-SSA validation gave median NRMSE values of 3.46% for PST and 3.63% for GS-ST, with lower PST error in 28 of the 30 matched comparisons (Figure 5D). The gain therefore combines faster optimization and better prediction across initial kinetic-parameter values and stochastic realizations under the reduced training ensemble.

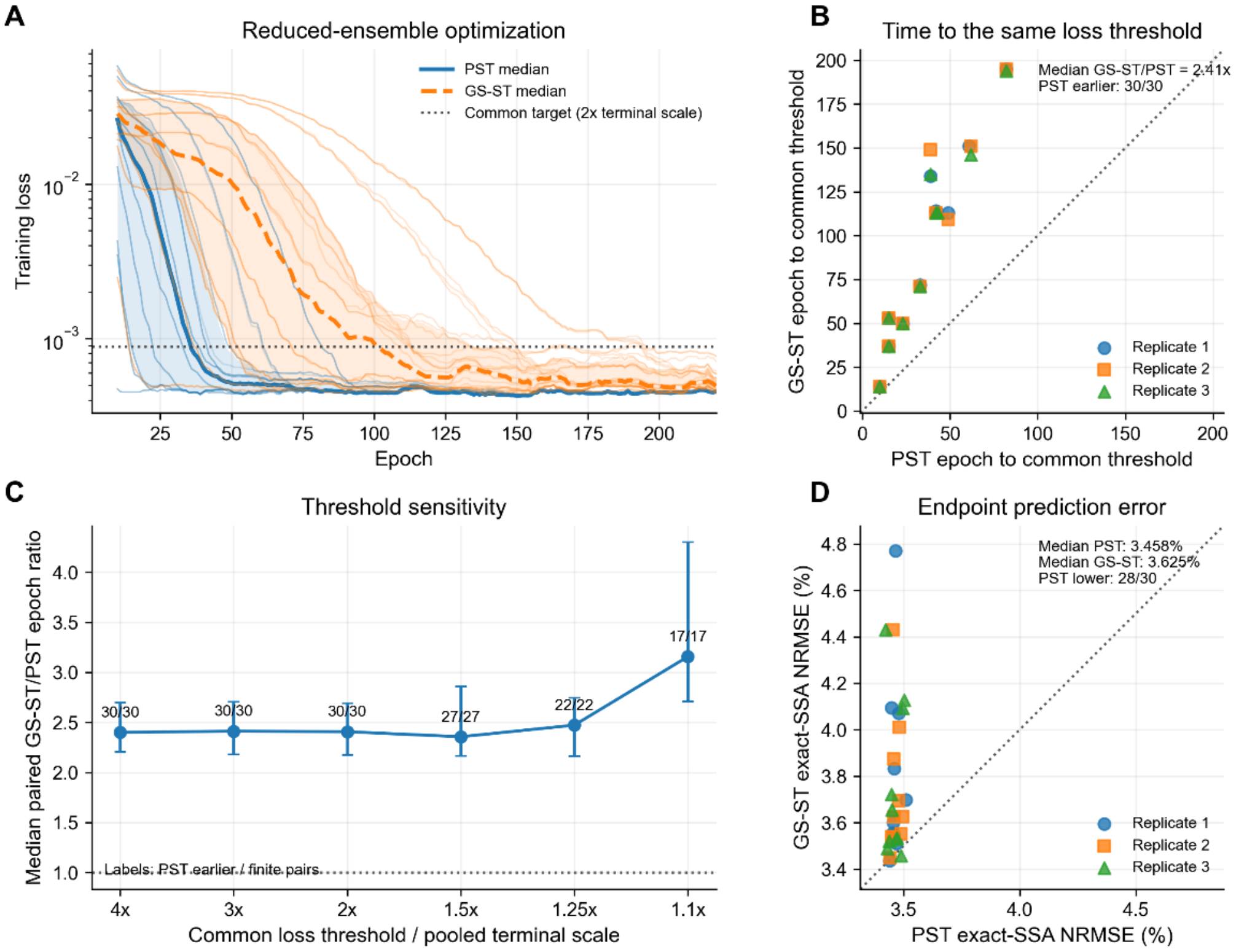

**Figure 5. Reduced-ensemble ion-channel inference.** Ion-channel kinetics were inferred using 10,000 training trajectories per epoch across ten sets of initial kinetic-parameter values and three stochastic seed blocks. **A,** Median-smoothed training losses for PST and GS-ST; the horizontal dotted line marks the common low-loss target. **B,** Paired epochs required to reach the common target; the diagonal denotes equal convergence epoch. **C,** Median paired GS-ST/PST convergence-epoch ratio across increasingly stringent common loss targets; error bars show interquartile ranges. Numbers indicate comparisons in which PST was earlier divided by comparisons in which both methods reached the target within 400 epochs. **D,** Independent exact-SSA validation NRMSE for the inferred models; the diagonal denotes equal error. In B and D, points above the diagonal favor PST.

## Accurate recovery across 50 repressilator tasks

The released repressilator model and task specification of Burger et al. (34) provide 50 externally specified inference problems. We retained the released stochastic parameterization, reference–initialization pairs, and task-construction logic, and compared PST and GS-ST using a common TensorFlow implementation adapted for this comparison. The main analysis uses relative mean-squared error rather than the log-of-ensemble-mean objective of the original study; the effect of this change is examined directly in the controlled objective comparison in SI Appendix, Table S1 and Figure S1. PST and GS-ST share the simulator, objective, optimizer, temporal readout, trajectory budget, and stopping procedure. All repressilator PST and GS-ST results reported in this section were generated under this common TensorFlow protocol.

For a representative task, both inferred parameter pairs reproduce the reference ensemble means at the fixed observation times (Figure 6A). Across the 50-task suite, both PST and GS-ST achieved two-parameter MAPE below 0.6% in 47 of 50 tasks and below 5.8% in all 50 tasks (Figure 6B). Median task-wise MAPE was 0.135% for PST and 0.137% for GS-ST, with corresponding mean values of 0.381% and 0.378%. The inferred production rates and dissociation constants closely track their reference values across the complete suite (Figure 6C,D), and independent exact-SSA validation produces nearly identical median relative mean-squared errors. The exploratory analysis by initial $K_d$ is reported in SI Appendix, Supplementary Analyses. The same temperature-free reaction-selection rule therefore achieves accurate parameter recovery across all 50 tasks.

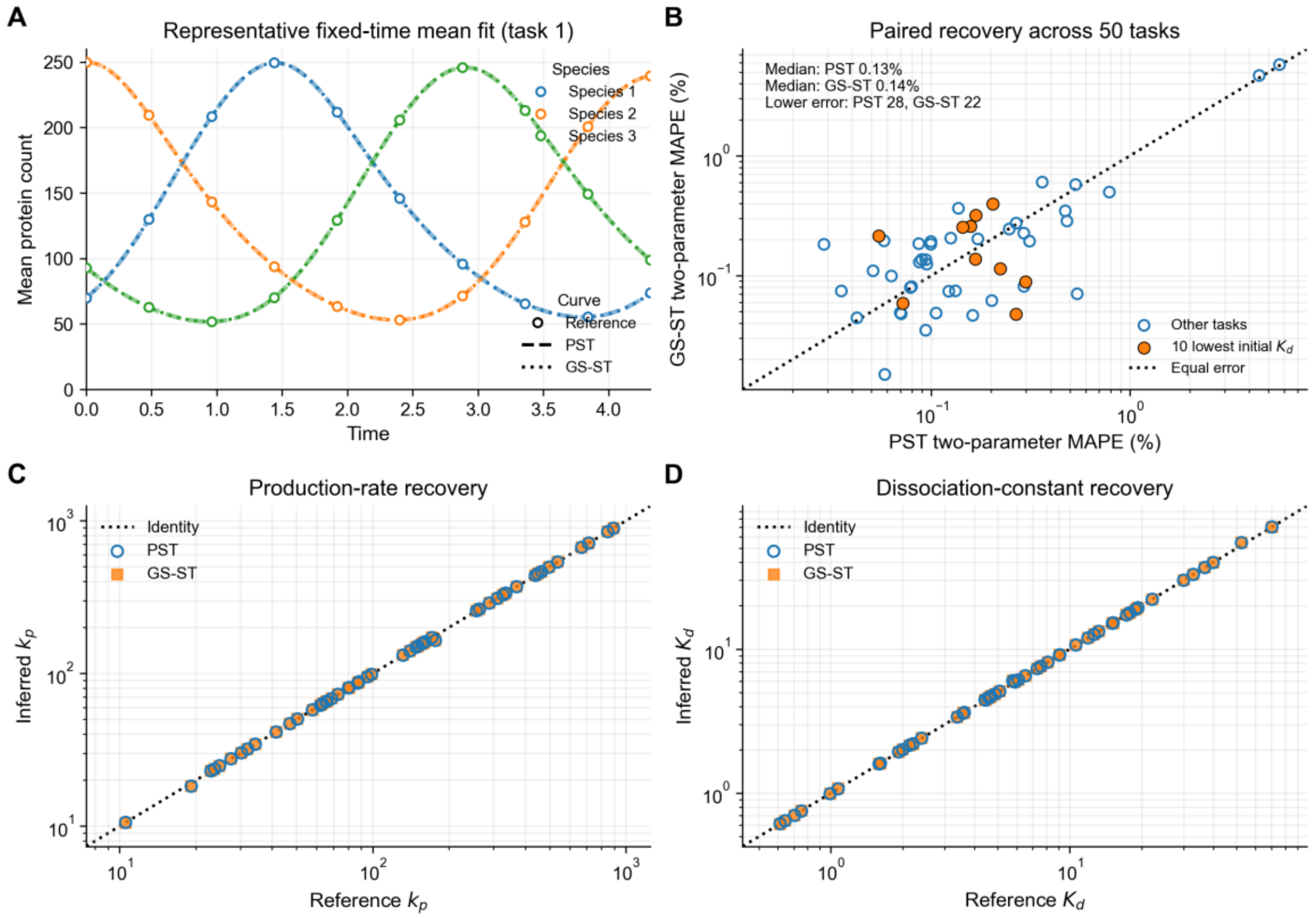

**Figure 6. PST on the 50-task stochastic repressilator benchmark.** The released stochastic parameterization and reference–initialization pairs were evaluated under the common adapted protocol described in Materials and Methods. All plotted PST and GS-ST results were generated under this common protocol. **A**, Representative fixed-time ensemble-mean trajectories for task 1. Open markers show the reference mean estimated from exact SSA trajectories; dashed and dotted curves show independent exact-SSA validation means at the PST- and GS-ST-inferred parameter pairs. **B**, Paired task-wise two-parameter MAPE for PST and GS-ST; the diagonal denotes equal error, and filled symbols identify the highlighted low-$K_d$ subset. **C**, Inferred production rate $k_p$ versus its reference value. **D**, Inferred dissociation constant $K_d$ versus its reference value. Dotted identity lines in **C** and **D** denote exact recovery. Both methods achieve two-parameter MAPE below 0.6% in 47 of 50 tasks and below 5.8% in all 50 tasks.

### Exact-event training at deep-learning parameter scale

PST trains 203,796 parameters with exact discrete events from initialization in the stochastic gene-regulatory MNIST classifier introduced in Ref. (18). PST is active from epoch 0, and every training forward pass uses exact hard categorical reaction selection, with no reaction-selection relaxation temperature or soft-forward warm-up. The earlier GS-ST implementation of the same architecture used soft-forward Gumbel-Softmax sampling during the first 16 of 80 training epochs before switching to hard-forward straight-through training (18). The same propensity-based rule therefore supports large-scale learning without either a relaxed training phase or temperature annealing. Test accuracy improves rapidly through approximately epoch 20 and more gradually thereafter during the 40-epoch PST run (SI Appendix, Figure S2B). The final single-pass test accuracy is 97.43%. The highest observed accuracy is 98.22%, using four Monte Carlo realizations and averaging over the terminal 64% of the trajectory.

## Discussion

We have shown that exact stochastic reaction probabilities support effective gradient-based learning directly. PST differentiates normalized reaction propensities in the backward pass, preserving exact event selection and exponential waiting times while eliminating the need for auxiliary Gumbel perturbations and reaction-selection temperature tuning. This construction yields the exact one-step conditional-mean sensitivity and admits an explicit analysis of its composition along nonlinear trajectories.

Composed along a trajectory, PST replaces the exact finite differences between reachable downstream states by directional derivatives propagated through the backward computational graph. The discrepancy vanishes when the relevant continuation value is affine. When downstream curvature and reaction-probability sensitivities remain bounded, the discrepancy decreases as reaction jumps become small relative to the scale over which downstream values vary, connecting PST to classical small-jump expansions of Markov jump processes without replacing the underlying Markov jump process by a diffusion (27–30, 35). At leading order, the discrepancy separates downstream curvature from the parameter sensitivity of the conditional jump covariance. The latter is determined by the network and parameterization, whereas loss and readout design provide directly controllable influences on downstream curvature. For weighted least-squares objectives that compare population means, any parameter setting for which the model population mean exactly matches the target mean is stationary under both the exact and PST update fields.

The empirical gains are both computational and statistical. In reversible dimerization, PST lowers the mean parameter MAPE from 0.093% to 0.060%. In the genetic oscillator, it reduces median run-wise MAPE from 2.58% to 1.70% at the matched 1000-epoch horizon and reaches sustained error below 3% about threefold sooner. In reduced-ensemble ion-channel inference, PST converges earlier in all 30 matched comparisons, with a median paired GS-ST/PST epoch ratio of 2.41. Independent exact-SSA validation gives lower PST prediction error in 28 of 30 comparisons and a median NRMSE of 3.46% versus 3.63%. These gains are obtained without a reaction-selection temperature or annealing schedule. The repressilator suite extends accurate recovery to all 50 tasks, and the large stochastic classifier demonstrates exact-event training from initialization without a soft-forward warm-up. Together, the results show that removing the backward relaxation improves learning performance while simplifying optimization.

PST complements existing CTMC sensitivity approaches, including unbiased estimators (22), coupled finite differences (23), score-function estimators (15), and pathwise methods (24, 25). The exact local reaction-selection gradient has a score-function representation with the current-state continuation value as a natural baseline. PST replaces the finite-difference advantage by a probability-weighted directional derivative, yielding a differentiable local rule that is exact for affine continuation values and has an explicit curvature-dependent error bound. At the categorical level, PST is closely related to marginal-based straight-through estimation (20), but the affine SSA update gives the SSA construction its exact one-step conditional-mean property.

PST turns probabilities already required for exact simulation into learning signals. Its conditional-mean foundation and curvature–covariance analysis connect this simple construction to stochastic-process theory, while the inference and classifier results demonstrate its reach from experimental kinetic modeling to deep-learning-scale optimization. By eliminating reaction-selection temperature tuning while accelerating inference and enabling exact-event training from initialization, PST provides a direct route to learning mechanistic stochastic models without smoothing away their discrete dynamics.

# Materials and Methods

## Estimator implementation and comparison design

For the kinetic-inference benchmarks, all trainable kinetic parameters are represented in log space to enforce positivity. The PST calculations used Python 3.11.13 and TensorFlow 2.20.0. PST uses direct hard categorical sampling and differentiates the normalized propensities in the backward pass, whereas GS-ST uses a hard Gumbel-Max reaction with a temperature-dependent Gumbel-Softmax backward sample (36, 37, 18). Within each of the four inference benchmarks, PST and GS-ST use the same state model, objective, optimizer, learning-rate schedule, ensemble size, initialization protocol, and simulation horizon unless explicitly stated otherwise.

The oscillator comparison uses matched initialization multipliers, target trajectories, and nominal run seeds across methods, providing a closely aligned protocol for evaluating optimization behavior. For dimerization and the oscillator, GS-ST reaction-selection temperatures were retained at the values established in Ref. (18). The ion-channel and repressilator settings are specified separately below; PST has no reaction-selection temperature. The ion-channel comparison in Figure 4 uses one optimization realization per method; the robustness analysis in Figure 5 uses ten sets of initial kinetic-parameter values in each of three stochastic seed blocks. The repressilator benchmark evaluates one seeded realization per method for each task under the common adapted protocol. The classifier complements these inference benchmarks by testing hard-forward PST from initialization at deep-learning scale. Full framework-specific code, tensor conventions, random-stream details, and absorbing-state handling are provided in SI Appendix, SI Materials and Methods.

## Statistical analysis

For a parameter vector with inferred components, parameter mean absolute percentage error (MAPE) is given by $\mathrm{MAPE}(\hat{\boldsymbol{\theta}}, \boldsymbol{\theta}^*) = \frac{100}{Q}\sum_{q=1}^{Q}\left|\frac{\hat{\theta}_q-\theta_q^*}{\theta_q^*}\right|$. For dimerization, MAPE is computed separately for each of the eight kinetic conditions and then averaged across conditions. Repressilator results use task-wise two-parameter MAPE, summarized by the median across the 50 tasks.

For each oscillator run, each inferred parameter is summarized by its geometric mean over the terminal 25% of the selected training horizon, after which parameter MAPE is computed. Run-level results are summarized by medians and interquartile ranges, with 95% percentile bootstrap intervals obtained by resampling runs. Threshold analyses use a trailing 50-epoch geometric mean of parameter MAPE and require the threshold to remain satisfied for 50 consecutive epochs.

Ion-channel fits are evaluated by root-mean-square error (RMSE), normalized root-mean-square error (NRMSE; RMSE divided by the experimental mean-trace range), and $R^2$. The descriptive stabilization epoch is the first completed epoch after which a trailing 10-epoch arithmetic mean remains within a specified relative band of the final 50-epoch mean loss. Full aggregation orders, smoothing windows, bootstrap procedures, and sensitivity analyses are given in SI Appendix.

For Figure 5, training losses are smoothed with a trailing 10-epoch median. For each run, the terminal loss is defined as the median loss over the final 50 epochs, and the common terminal loss scale is the median of these run-specific values pooled across PST and GS-ST. The primary common target is twice this terminal loss scale. The convergence epoch is the first epoch at which the smoothed loss remains below the common target for 10 consecutive epochs. Sensitivity to the target choice is assessed using additional common targets equal to 4, 3, 1.5, 1.25, and 1.1 times the terminal loss scale; sensitivity to the persistence requirement is assessed using 5, 10, and 25 consecutive epochs at the primary target. For each target, the paired GS-ST/PST convergence-epoch ratio is calculated only when both estimators reach the target within the 400-epoch training horizon. The 30 matched comparisons comprise ten sets of initial kinetic-parameter values evaluated within each of three stochastic seed blocks and are treated descriptively rather than as 30 independent stochastic replicates.

## Reversible dimerization

The benchmark is the reversible heterodimerization $A + B \underset{k_2}{\overset{k_1}{\rightleftharpoons}} C$, with propensities $a_1 = k_1 AB$ and $a_2 = k_2 C$. The initial state is $(A, B, C) = (100, 90, 0)$, the generating forward rate is $k_1 = 0.01$, and the reverse rate takes eight values from 0.01 to 1.28. For each condition, the target is generated from 100,000 exact SSA trajectories of 250 reactions. Training uses 100,000 model trajectories per epoch for 250 epochs and minimizes squared error between target and model ensemble-mean time courses after mapping each exact trajectory to a common physical-time grid. Individual trajectories are interpolated before averaging. The full time-grid construction, interpolation conventions, objective, optimizer, gradient clipping, and learning-rate schedule are specified in SI Appendix.

## Genetic oscillator

The oscillator is the nine-species, sixteen-reaction network of Vilar et al. (31). Five parameters, $(\beta_A, \beta_R, \delta_A, \delta_R, \gamma_C)$, are inferred simultaneously. A 2,400,000-reaction exact SSA reference trajectory defines 262,144 starting states and associated 150-reaction target segments. Each epoch samples 8192 starting states and generates 25 model trajectories per state. The objective compares segment-level rate statistics for the activator, repressor, and activator–repressor complex.

The comparison uses ten newly generated 3000-epoch PST runs and the ten-run GS-ST result set from Ref. (18). Both methods use the same log-uniform initialization multipliers, target construction, optimizer, learning-rate schedule, and true-value-centered parameter bounds. The GS-ST comparison uses the fixed reaction-selection temperature $T = 3 \times 10^{-6}$ set from Ref. (18). Complete objective definitions, seed schedules, numerical guards, optimization settings, and sampling details are given in SI Appendix.

## Ion-channel inference

The experimental model has two channels and three states, $C \underset{k_{\text{close}}}{\overset{k_{\text{open}}}{\rightleftharpoons}} O \overset{k_{\text{inact}}}{\rightarrow} I$, with $I$ absorbing over the recording window. The data are 100 idealized HEK293 cell-attached patch-clamp sweeps at $+40$ mV from the publicly available $Na_V1.5$ dataset of Selimi et al. (32, 33). For the Figure 4 comparison, training uses 262,144 model trajectories per epoch, 20 SSA event slots per trajectory, and 400 epochs. The objective compares the experimental mean open-channel count with exact overlap integrals of the model's piecewise-constant open-channel trajectories over the experimental time bins.

All three rates are optimized in log space from $0.25\ \mathrm{ms}^{-1}$ with RMSprop. The GS-ST reaction-selection temperature is annealed exponentially from 0.1 to 0.001, with the comparator based on the earlier ion-channel implementation in Ref. (18); the corresponding learning rate decreases from 0.05 to 0.0005. PST has no reaction-selection temperature and uses the same learning-rate schedule. Final fits use 30,000 independent hard-forward SSA trajectories. The bin integration, event-time conventions, continuation rules, and optimizer schedule are described in SI Appendix.

For the Figure 5 robustness analysis, the same ion-channel model and objective are optimized for 400 epochs using 10,000 model trajectories per epoch and 40 SSA event slots per trajectory. Ten sets of initial kinetic-parameter values are evaluated in each of three stochastic seed blocks. The original $(0.25,\ 0.25,\ 0.25)\ \mathrm{ms}^{-1}$ initialization is retained as one starting point, and the remaining starting rates span $0.2$–$1.0\ \mathrm{ms}^{-1}$ using a fixed log-space Latin-hypercube design. PST and GS-ST use the same matched RMSprop learning-rate schedule as in Figure 4, from 0.05 to 0.0005. GS-ST likewise retains the same exponentially annealed reaction-selection temperature schedule, from 0.1 to 0.001, whereas PST has no reaction-selection temperature. Endpoint fits use 300,000 independent hard-forward SSA trajectories.

## Repressilator inference

The benchmark is a reduced three-protein stochastic repressilator (38), with the stochastic parameterization and 50 reference–initialization pairs adapted from Burger et al. (34). The inferred parameters are the production rate $k_p$ and dissociation constant $K_d$. Each task uses 50,000 exact trajectories to construct the target mean, 50,000 trajectories per stochastic loss and gradient evaluation, and 50,000 independent trajectories for final validation.

PST and GS-ST use the same exact SSA forward process, relative mean-squared-error objective, temporal straight-through readout, optimizer, stopping rule, and refinement stage. They differ in the backward treatment of categorical reaction selection: PST differentiates the normalized propensities, whereas GS-ST differentiates a Gumbel-Softmax relaxed sample with temperature of $0.30$. One optimization realization is performed for each method and task. The comparison uses a common TensorFlow implementation and evaluates the two backward estimators within the shared temporal-readout approximation. Complete propensities, initial-state and observation-time construction, temporal readout, sensitivity propagation, optimization protocol, and objective-control experiment are provided in SI Appendix.

## Deep-learning-scale classifier

The stochastic gene-regulatory classifier maps 784 input transcription-factor concentrations through 256 hidden genes to 10 output genes and contains 203,796 trainable parameters (18). The PST calculation uses the published architecture, MNIST data split, stochastic reaction-network dynamics, categorical cross-entropy objective, RMSprop optimizer, stochastic weight averaging, and post-training Monte Carlo and temporal averaging. PST is used from the first epoch throughout the 40-epoch training run, with exact hard categorical reaction selection in every forward pass and no reaction-selection relaxation temperature or soft-forward warm-up. The classifier serves as a PST scale demonstration using a different training schedule from the earlier GS-ST run. Full training and evaluation details are given in SI Appendix.

## Acknowledgments

J.M.G.V. acknowledges support from Ministerio de Ciencia, Innovación y Universidades (Grant PID2024-160016NB-I00 funded by MICIU/AEI/10.13039/501100011033 and by ERDF/EU).

## Author Contributions

J.M.G.V. and L.S. contributed equally to this work. Both authors designed the research, developed the methodology, performed the research, analyzed the data, and wrote and revised the manuscript.

## Competing Interests

The authors declare no competing interests.

## Data, Materials, and Software Availability

Code and data generated for this study are available at `https://github.com/CTMCDL/PST`. These materials include the PST implementations and analysis code; the PST and GS-ST oscillator result files used in the comparison; the TensorFlow repressilator adaptation and results used for Figure 6 and the supplementary objective-control analysis; the ion-channel training histories, multi-initial-value analysis, and convergence analysis used for Figures 4 and 5; and the PST MNIST notebook, saved training history and weights, Monte Carlo–temporal-averaging results, and figure source files.

# SI Appendix

## Propensity Straight-Through Gradients for Discrete Stochastic Systems

Jose M. G. Vilar[1,2,*] and Leonor Saiz[3,*]
[1]Biofisika Institute (CSIC, UPV/EHU), University of the Basque Country, P.O. Box 644, 48080 Bilbao, Spain
[2]IKERBASQUE, Basque Foundation for Science, 48011 Bilbao, Spain
[3]Department of Biomedical Engineering, University of California, 451 East Health Sciences Drive, Davis, CA 95616, USA

[*] To whom correspondence should be addressed: j.vilar@ikerbasque.org or lsaiz@ucdavis.edu

### Contents

## SI Materials and Methods

### PST implementation and tensor conventions

For a nonabsorbing state $\mathbf{x}$, normalized propensities are

$$\pi_j(\mathbf{x};\boldsymbol{\theta}) = \frac{a_j(\mathbf{x};\boldsymbol{\theta})}{\sum_{\ell=1}^{M} a_\ell(\mathbf{x};\boldsymbol{\theta})}. \tag{S1}$$

The exact SSA samples a hard categorical reaction $J \sim \mathrm{Categorical}(\boldsymbol{\pi})$, forms $\mathbf{y} = \mathbf{e}_J$, and updates the state as $\mathbf{x}^+ = \mathbf{x} + \mathbf{S}\mathbf{y}$. PST leaves this forward reaction selection unchanged and replaces only the backward derivative through the reaction indicator:

$$\mathbf{y}_{\mathrm{PST}} = \boldsymbol{\pi} + \mathrm{stopgrad}(\mathbf{y} - \boldsymbol{\pi}). \tag{S2}$$

Numerically, $\mathbf{y}_{\mathrm{PST}} = \mathbf{y}$, while automatic differentiation gives $\nabla_{\boldsymbol{\theta}}\mathbf{y}_{\mathrm{PST}} = \nabla_{\boldsymbol{\theta}}\boldsymbol{\pi}$. We use numerator-layout Jacobians. Thus, for $\boldsymbol{\pi} \in \mathbb{R}^M$ and $\boldsymbol{\theta} \in \mathbb{R}^P$,

$$\nabla_{\boldsymbol{\theta}}\boldsymbol{\pi} \equiv \frac{\partial \boldsymbol{\pi}}{\partial \boldsymbol{\theta}} \in \mathbb{R}^{M\times P}, \tag{S3}$$

so that $\mathbf{S}\,\partial\boldsymbol{\pi}/\,\partial\boldsymbol{\theta} \in \mathbb{R}^{N\times P}$ when the stoichiometric vectors are columns of $\mathbf{S}$. In the TensorFlow implementation below, the batch-oriented multiplication uses the corresponding row layout, with the one-hot reaction on the left and the reaction-by-species stoichiometry matrix on the right. Gradients of scalar functions are represented as column vectors.

```
propensity = propensity_fn(state, parameters)
probability = propensity / tf.reduce_sum(propensity, axis=-1, keepdims=True)
reaction = tf.random.stateless_categorical(tf.math.log(propensity), 1, seed)
y_hard = tf.one_hot(tf.squeeze(reaction, -1), depth=tf.shape(propensity)[-1])
y_pst = probability + tf.stop_gradient(y_hard - probability)
state_increment = tf.matmul(y_pst, stoichiometry)
```

Hard categorical sampling uses log-propensities as logits. This representation is stable when propensities span a wide range and maps directly to optimized categorical-sampling kernels; subtracting the common $\log a_0$ is unnecessary because it does not change the categorical distribution. Zero propensities correspond to $-\infty$ logits. Absorbing rows with $a_0 = 0$ are masked before sampling, no reaction is selected, and the state remains unchanged. Leading batch dimensions may be flattened for the categorical kernel.

At a fixed state, the waiting time is sampled by the exact reparameterization

$$\Delta t = -\frac{\log u}{a_0}, \qquad u \sim \mathrm{Uniform}(0,1). \tag{S4}$$

PST is not used for this direct derivative. The waiting-time qualification for later state-dependent event times is derived in Supplementary Theory.

## Common implementation and statistical summaries

All trainable kinetic parameters are represented in log space to enforce positivity. The PST reruns were executed with Python 3.11.13 and TensorFlow 2.20.0; the published GS-ST notebooks document the same software versions for the archived baseline (1). PST uses direct hard categorical sampling and Eq. (S2) in the backward pass. GS-ST uses an exact Gumbel-Max hard reaction with a Gumbel-Softmax relaxed backward sample. Within each inference benchmark, the comparisons retain the same state model, objective, optimizer, learning-rate schedule, ensemble size, initialization protocol, and simulation horizon. The classifier scale demonstration uses the distinct training schedule detailed below.

In the oscillator, the nominal run seeds, initialization multipliers, and inference target trajectory are matched across methods.

For a parameter vector with $Q$ inferred components, the parameter mean absolute percentage error (MAPE) is

$$\mathrm{MAPE}(\hat{\boldsymbol{\theta}}, \boldsymbol{\theta}^*) = \frac{100}{Q} \sum_{q=1}^{Q} \left| \frac{\hat{\theta}_q - \theta_q^*}{\theta_q^*} \right|. \tag{S5}$$

The condition-level MAPE is Eq. (S5) for one dimerization condition; the reported dimerization mean is the average across the eight values of $k_2$. For the repressilator benchmark, Eq. (S5) is evaluated for the two inferred parameters in each task, and the primary summary is the median across the 50 task-wise MAPEs.

Three temporal summaries are used for the oscillator. The epoch-history figure displays the epoch-indexed parameter-MAPE histories. Threshold times use a trailing 50-epoch geometric mean and require the smoothed value to remain below threshold for 50 consecutive epochs. Horizon-specific parameter recovery uses a separate summary: for each run and horizon, every positive parameter is aggregated geometrically over the terminal 25% of epochs, and Eq. (S5) is applied to that window-aggregated vector. The 1000- and 3000-epoch summaries therefore use windows of 250 and 750 epochs, respectively. The primary run-wise MAPE is the median of the ten run-level values. The interquartile range is the 25th–75th percentile range, and the reported 95% confidence intervals are percentile bootstrap intervals obtained by resampling runs.

The global pooled-vector MAPE uses the same terminal 25% window and log-space geometric aggregation but changes the aggregation order: parameter samples are pooled across runs before a single geometric-mean parameter vector is formed. The run-wise statistic instead forms one geometric-mean vector per run and then summarizes the resulting run-level MAPEs. At 3000 epochs, the pooled-vector MAPE is 1.69% for PST and 1.23% for GS-ST.

For ion-channel fits, root-mean-square error (RMSE) is computed between the experimental and simulated mean traces. Normalized RMSE (NRMSE) is RMSE divided by the range of the experimental mean trace. The coefficient of determination $R^2$ uses the usual residual and total sums of squares. The descriptive loss-stabilization epoch is the first completed epoch after which a trailing 10-epoch arithmetic mean of the training loss remains no more than $5\%$ above the mean loss over the final 50 epochs. Sensitivity to $20\%$, $10\%$, and $2\%$ bands is reported in Table S3.

Oscillator convergence is evaluated from a trailing 50-epoch geometric mean of instantaneous parameter MAPE, followed by a requirement that the smoothed value remain below $3\%$ for 50 consecutive epochs. All ten runs in each arm satisfy this criterion by epoch 3000. The ratio of median threshold epochs is accompanied by a bootstrap interval. The GS-ST/PST ratio of median crossing epochs is approximately 3.0. Under the common implementation described in the main Results and Section S8, equal per-epoch cost gives approximately one-third of the optimization time for PST; no-greater PST per-epoch cost is sufficient for the corresponding time bound.

## Reversible dimerization protocol

The benchmark is a reversible heterodimerization,

$$A + B \underset{k_2}{\overset{k_1}{\rightleftharpoons}} C,$$

with propensities $a_1 = k_1 AB$ and $a_2 = k_2 C$. The initial state is $(A, B, C) = (100, 90, 0)$. The true forward rate is fixed at $k_1 = 0.01$, and the reverse rate takes the eight values

$$k_2 \in \{0.01, 0.02, 0.04, 0.08, 0.16, 0.32, 0.64, 1.28\}.$$

For each condition, the target is generated from 100,000 exact SSA trajectories of 250 reactions. Each trajectory is mapped to a common physical-time grid ending at the minimum terminal time across the target ensemble, and the species means are computed on that grid. For event times $\tau_0 < \cdots < \tau_L$ and one species value $z_\ell$ stored at each event time, the readout used in the previous comparison uses linear interpolation,

$$\mathcal{I}_z(t) = z_\ell + \frac{t - \tau_\ell}{\tau_{\ell+1} - \tau_\ell}(z_{\ell+1} - z_\ell), \qquad \tau_\ell \le t < \tau_{\ell+1}. \tag{S6}$$

The bracketing interval is selected with a right-sided search. Therefore, a query exactly at an event time uses the post-event state stored at that time. A $10^{-8}$ time guard enforces strict ordering in degenerate numerical cases. The search indices are treated as fixed by automatic differentiation; derivatives propagate through the selected event times, state values, and interpolation fraction. The target-grid endpoint ensures that every target trajectory brackets the complete grid. If a 250-event model trajectory ends before that endpoint during training, the implementation uses the last two simulated points and linearly extrapolates Eq. (S6) beyond the final event time. Thus, the event states and waiting times are generated by exact hard SSA, whereas the fixed-time readout is a differentiable linear interpolation of those states.

Training uses 100,000 model trajectories per epoch, one initial condition per batch, and 250 epochs. Let $K_{\mathrm{dim}}$ denote the number of points on the common physical-time grid. The loss is

$$\mathcal{L}_{\mathrm{dim}} = \frac{1}{K_{\mathrm{dim}}} \sum_{i=1}^{K_{\mathrm{dim}}} \left[ (\hat{A}_i - A_i^*)^2 + (\hat{B}_i - B_i^*)^2 + (\hat{C}_i - C_i^*)^2 \right], \tag{S7}$$

where the hats denote model ensemble means and the stars denote target ensemble means. Individual model trajectories are interpolated before averaging. The parameters are initialized at $k_1 = 0.125$ and $k_2 = 0.025$ and optimized with RMSprop. Gradients are clipped to a global norm of 5. The schedule used in Ref. (1) defines $q_e$ exponentially from 1 to 0.001 and sets the learning rate to $0.2q_e^{1.5}$. For GS-ST, $q_e$ is also the relaxation temperature; for PST, it only parameterizes the matched learning-rate schedule.

## Genetic oscillator protocol

The oscillator is the nine-species, sixteen-reaction network of Vilar et al. (2). Five parameters are inferred simultaneously:

$$(\beta_A, \beta_R, \delta_A, \delta_R, \gamma_C) = (50, 5, 1, 0.2, 2)$$

in the generating model. Each run begins from $(1, 0, 1, 0, 0, 0, 0, 0, 0)$. A 2,400,000-reaction exact reference trajectory is used to construct a map of 262,144 starting states and associated 150-reaction target segments. At each epoch, 8192 starting states are sampled and 25 model trajectories are generated per starting state, for 204,800 trajectories in the gradient calculation.

For species $Z \in \{A, R, C\}$, let $\overline{Z}_{100}$ and $\bar{t}_{100}$ denote the arithmetic means of the species value and event time, respectively, over the final 100 events of a segment. The segment-level rate statistic is

$$r_Z = \frac{\overline{Z}_{100} - Z_0}{\bar{t}_{100} - t_0 + 10^{-9}}. \tag{S8}$$

The objective is

$$\mathcal{L}_{\mathrm{osc}} = \mathrm{MSE}(r_A, r_A^*) + \mathrm{MSE}(r_R, r_R^*) + \mathrm{MSE}(r_C, r_C^*). \tag{S9}$$

The $10^{-9}$ term in Eq. (S8) is a numerical division guard. The optional terminal-time term is disabled ($\lambda_t = 0$). The objective is event-segment based rather than a fixed-observation-time loss, although it remains a nonlinear multistep objective.

The comparison uses ten newly generated PST runs and the archived ten-run GS-ST result set from the published Vilar–Saiz study (1), each with 3000 epochs. Both use the same log-uniform initialization multipliers generated from base seed 43 and the same nominal run-seed schedule, $43 + 97377r$ for run index $r = 0, \ldots, 9$. For each matched run, both methods use the same 2,400,000-reaction reference trajectory and the same stratified target segments for inference.

RMSprop starts at learning rate 0.1. Every 25 epochs, the learning rate is updated according to

$$\eta \leftarrow (\eta - 0.001)\, 0.9 + 0.001. \tag{S10}$$

Gradients are clipped to global Euclidean norm 1.0. After each update, every inferred log-parameter is projected to within $\log 10$ of its generating value, restricting each inferred rate to the interval $[0.1\theta_p^*, 10\theta_p^*]$. This true-value-centered constraint is retained from the published synthetic benchmark and is applied to both methods.

PST uses `tf.random.stateless_categorical` for the hard reaction. The archived GS-ST runs use the fixed temperature used in Ref. (1) $T = 3 \times 10^{-6}$, which was characterized for this system.

## Ion-channel protocol

The experimental model has two channels and three states,

$$C \underset{k_{\text{close}}}{\overset{k_{\text{open}}}{\rightleftharpoons}} O \overset{k_{\text{inact}}}{\rightarrow} I,$$

with $I$ absorbing over the recording window. The mass-action propensities are

$$a_{\text{open}} = k_{\text{open}} n_C, \qquad a_{\text{close}} = k_{\text{close}} n_O, \qquad a_{\text{inact}} = k_{\text{inact}} n_O.$$

All simulations begin from $(n_C, n_O, n_I) = (2, 0, 0)$. The fitted data are 100 idealized HEK293 cell-attached patch-clamp sweeps at $+40$ mV from the publicly archived $Na_V1.5$ dataset of Selimi et al. (3, 4). The observed value is the number of open channels, $n_O \in \{0, 1, 2\}$.

Training uses 262,144 model trajectories per epoch, 20 SSA event slots per trajectory, and 400 epochs. Let $t_1 < \cdots < t_{K_{\text{ion}}}$ denote the experimental time grid, and let $\tau_\ell$ and $n_{O,\ell}$ denote the stored event times and the right-continuous open-channel state on $[\tau_\ell, \tau_{\ell+1})$. Define the model mean open-channel count in experimental bin $i$ as the exact overlap integral of this piecewise-constant path,

$$\widehat{m}_i = \frac{1}{t_{i+1} - t_i} \sum_{\ell=0}^{L} n_{O,\ell} \left[\min(t_{i+1}, \tau_{\ell+1}) - \max(t_i, \tau_\ell)\right]_+, \qquad \tau_{L+1} = +\infty. \tag{S11}$$

The last stored state is continued after the final simulated event. The implementation evaluates the equivalent difference of cumulative areas. A right-sided interval search assigns a time exactly coinciding with an event to the post-event interval; this boundary convention does not alter the integral but fixes the automatic-differentiation branch. A $10^{-8}$ time ramp enforces strict ordering, and no derivative is taken through the discrete search indices. Gradients propagate through the selected event times, states, overlap lengths, and cumulative areas.

Let $m_i^*$ denote the experimental comparison value for bin $i$, defined as the average of the two endpoint means. The objective is

$$\mathcal{L}_{\text{ion}} = \frac{1}{K_{\text{ion}} - 1} \sum_{i=1}^{K_{\text{ion}}-1} (\hat{m}_i - m_i^*)^2. \tag{S12}$$

All three rates are initialized at $0.25\ \text{ms}^{-1}$ and optimized in log space with RMSprop. The comparison uses the following ion-channel GS-ST schedule: $q_e$ decreases exponentially from 0.1 to 0.001 and the learning rate is $\eta_e = 0.05(q_e/0.1)$. For GS-ST, $q_e$ is the reaction-selection temperature; for PST, the estimator ignores $q_e$ and the same numerical schedule controls only the learning rate. No additional GS-ST temperature search is performed. Final fits are evaluated with 30,000 independent hard-forward SSA trajectories, without any backward relaxation.

For the reduced-ensemble robustness analysis in main-text Figure 5, the same ion-channel model and objective are optimized for 400 epochs using 10,000 trajectories per epoch and 40 SSA event slots per trajectory. Ten sets of initial kinetic-parameter values are evaluated in each of three stochastic seed blocks; the original (0.25, 0.25, 0.25) $\text{ms}^{-1}$ initialization is retained, and the remaining starting rates span 0.2-1.0 $\text{ms}^{-1}$ using a fixed log-space Latin-hypercube design. PST and GS-ST retain the Figure 4 learning-rate schedule, and GS-ST retains the same exponentially annealed reaction-selection temperature schedule from 0.1 to 0.001. Final validation uses 300,000 independent exact SSA trajectories. Training losses are smoothed with a trailing 10-epoch median; the common terminal loss scale is the median of the run-specific final-50-epoch median losses pooled across PST and GS-ST, and the primary convergence target is twice this value. The convergence epoch is the first epoch for which the smoothed loss remains below the target for 10 consecutive epochs. Sensitivity is assessed at targets of 4, 3, 1.5, 1.25, and 1.1 times the terminal loss scale and persistence requirements of 5, 10, and 25 epochs; paired GS-ST/PST ratios are calculated only when both estimators reach the target within 400 epochs. The 30 matched comparisons comprise ten sets of initial kinetic-parameter values evaluated within each of three stochastic seeds.

## Repressilator inference protocol

The model is a reduced three-protein stochastic repressilator (5). Its stochastic parameterization, 50 reference–initialization parameter pairs, and task-construction logic were adapted from Burger et al. (6). The model contains three proteins $P_1, P_2, P_3$ and six reactions. With unit volume, unit degradation rate, and Hill coefficient $h = 3$, the cyclic production and degradation propensities are

$$\begin{aligned} a_{+1} &= \frac{k_p}{1 + (P_3/K_d)^h}, \qquad a_{-1} = P_1, \\ a_{+2} &= \frac{k_p}{1 + (P_1/K_d)^h}, \qquad a_{-2} = P_2, \\ a_{+3} &= \frac{k_p}{1 + (P_2/K_d)^h}, \qquad a_{-3} = P_3. \end{aligned} \tag{S13}$$

The inferred parameters are the production rate $k_p$ and dissociation constant $K_d$; $h$ remains fixed. The released task array supplies 50 reference pairs and 50 corresponding initial pairs. The reference values span $10 \le k_p \le 1000$ and $k_p/75 \le K_d \le k_p/3$. Following the released driver, each task's initial molecular state and observation horizon are constructed from the reference mean-field oscillator: the deterministic equations are integrated from $(1, 0, 0)$ to time 100, the last two peaks of $P_2$ define the period, and the rounded state at the last peak defines the stochastic initial condition, with any zero component replaced by one. Ten equally spaced observation times cover one period, leading to $K_{\mathrm{rep}} = 10$.

For each task, the target mean is estimated from 50,000 exact physical-time SSA trajectories. Training uses 50,000 trajectories per stochastic loss and gradient evaluation, and final validation uses 50,000 independent exact SSA trajectories. The hard forward trajectory is exact. PST differentiates the normalized propensities, whereas GS-ST uses a hard Gumbel-Max reaction with a Gumbel-Softmax backward temperature of $0.30$.

Both methods use the same fixed-time temporal straight-through readout: the hard forward indicator assigning an event increment to an observation time is retained, while its backward derivative is supplied by a logistic surrogate with temperature $0.05$. Burger et al. (6) instead replace the fixed-time Heaviside indicator with a sigmoid in the differentiable observable. For memory efficiency, the two log-parameter sensitivities are propagated explicitly through the physical-time event loop rather than by retaining a reverse-mode tape; this changes the implementation but not the PST or GS-ST surrogate definitions.

Let $\hat{\mu}_{is}$ and $\mu_{is}^{*}$ denote the model and target ensemble means for observation index $i$ and species $s$. The optimized objective is the relative mean-squared error

$$\mathcal{L}_{\mathrm{rep}} = \frac{1}{3K_{\mathrm{rep}}} \sum_{i=1}^{K_{\mathrm{rep}}} \sum_{s=1}^{3} \left( \frac{\hat{\mu}_{is} - \mu_{is}^{*}}{\mu_{is}^{*}} \right)^{2}, \tag{S14}$$

with exactly zero target means replaced numerically by $10^{-20}$. The two parameters are optimized in log space with RMSprop ($\eta = 0.1$, $\rho = 0.9$, zero momentum) and global gradient-norm clipping at 0.1. The nominal Stage-1 learning-rate rule updates every 200 evaluations according to $\eta \leftarrow (\eta - 0.001)0.9 + 0.001$. Stage 1 permits at most 1,000 stochastic evaluations and uses the released 50-evaluation loss-SNR stopping rule with threshold 0.01; the selected Stage-1 candidate is the lowest measured loss in the final rolling window. The adapted protocol then performs 50 refinement evaluations from this candidate using a fresh RMSprop optimizer at learning rate 0.01 and selects the lowest measured refinement loss, including the Stage-1 starting pair. No truth-dependent parameter clipping is applied. PST and GS-ST use separate stateless random-number streams, and one optimization realization is performed for each method and task.

As a controlled objective-sensitivity analysis, the complete protocol was repeated with the log-of-ensemble-mean objective used by Burger et al.; that supplementary run changes only the loss mode and is reported below.

# Supplementary Theory

## Detailed derivation of the local and multistep PST approximation

This note expands the derivation summarized in the main-text Results section "Local approximation and propagation across events". It isolates the contribution of one categorical reaction-selection node to the gradient of a downstream objective, compares the exact discrete contribution with the expected propensity straight-through (PST) contribution, and gives a worked two-step example.

## S1. Scope, notation, and the continuous extension

Fix a nonabsorbing state $\mathbf{x} \in \mathbb{Z}_{\geq 0}^{N}$. Reaction channel $j \in \{1, \dots, M\}$ is selected with normalized propensity

$$\pi_j(\mathbf{x}; \boldsymbol{\theta}) = \frac{a_j(\mathbf{x}; \boldsymbol{\theta})}{\sum_{\ell=1}^{M} a_\ell(\mathbf{x}; \boldsymbol{\theta})}, \tag{S15}$$

and changes the state by $\boldsymbol{\nu}_j$. Let

$$J \sim \text{Categorical}(\boldsymbol{\pi}), \qquad \mathbf{x}^+ = \mathbf{x} + \boldsymbol{\nu}_J. \tag{S16}$$

Let $V(\mathbf{z}; \boldsymbol{\theta})$ denote the continuation value, the expected downstream loss or the expected value of a scalar observable used to construct the loss, conditional on the post-reaction state being $\mathbf{z}$. It may be a post-reaction observable or include all later reactions, waiting times, and readout operations. On the lattice, the exact process requires $V$ only at reachable integer states. PST propagates ordinary derivatives with respect to state, requiring its backward rule to specify a differentiable extension of the downstream value to a neighborhood of those states. The finite differences below are intrinsic to the discrete process; the directional derivatives depend on that continuous extension. Two differentiable extensions can agree at every reachable lattice state while having different state derivatives at those states.

We use numerator-layout Jacobians: for $\boldsymbol{\pi} \in \mathbb{R}^M$ and $\boldsymbol{\theta} \in \mathbb{R}^P$, $\partial \boldsymbol{\pi} / \partial \boldsymbol{\theta} \in \mathbb{R}^{M \times P}$. The derivation is local to the current categorical node. Explicit parameter dependence of the continuation value contributes the direct term

$$\mathbf{g}_{\text{dir}} = \sum_{j=1}^{M} \pi_j(\mathbf{x}; \boldsymbol{\theta}) \nabla_{\boldsymbol{\theta}} V(\mathbf{x} + \boldsymbol{\nu}_j; \theta), \tag{S17}$$

which is carried recursively through the remaining computation. The comparison below isolates the additional contribution caused by changing the probabilities of the current reaction choice. The state $\mathbf{x}$ is held fixed for this local calculation; dependence of $\mathbf{x}$ on earlier reactions is handled by applying the same decomposition recursively.

## S2. Exact and PST reaction-selection contributions

The conditional expected downstream value after the current reaction is

$$F(\boldsymbol{\theta}) = \sum_{j=1}^{M} \pi_j(\mathbf{x}; \boldsymbol{\theta}) V(\mathbf{x} + \boldsymbol{\nu}_j; \boldsymbol{\theta}). \tag{S18}$$

Differentiating the probabilities while holding the post-reaction values fixed gives the exact reaction-selection contribution

$$\mathbf{g}_{\text{exact}} = \sum_{j=1}^{M} (\nabla_{\boldsymbol{\theta}} \pi_j)\, V(\mathbf{x} + \boldsymbol{\nu}_j). \tag{S19}$$

Because $\sum_j \nabla_{\boldsymbol{\theta}} \pi_j = \mathbf{0}$, any common scalar baseline may be subtracted. Choosing $V(\mathbf{x})$ gives

$$\mathbf{g}_{\text{exact}} = \sum_{j=1}^{M} (\nabla_{\boldsymbol{\theta}} \pi_j)\, [V(\mathbf{x} + \boldsymbol{\nu}_j) - V(\mathbf{x})]. \tag{S20}$$

Thus the exact contribution is weighted by the finite difference

$$\Delta_{\boldsymbol{\nu}_j} V(\mathbf{x}) = V(\mathbf{x} + \boldsymbol{\nu}_j) - V(\mathbf{x}). \tag{S21}$$

PST retains the hard forward state in Eq. (S16) but assigns the reaction indicator the backward derivative of its categorical mean. In matrix notation,

$$\frac{\partial \mathbf{x}_{\text{PST}}^{+}}{\partial \boldsymbol{\theta}} = \mathbf{S} \frac{\partial \boldsymbol{\pi}}{\partial \boldsymbol{\theta}}, \qquad \mathbf{S} = [\boldsymbol{\nu}_1 \quad \cdots \quad \boldsymbol{\nu}_M]. \tag{S22}$$

Conditional on the realized hard reaction $J$, the PST contribution propagated through the continuation value is

$$\left(\frac{\partial \mathbf{x}_{\text{PST}}^{+}}{\partial \boldsymbol{\theta}}\right)^{\mathsf{T}} \nabla_{\mathbf{x}} V(\mathbf{x} + \boldsymbol{\nu}_J). \tag{S23}$$

Define the probability-weighted post-reaction gradient

$$\overline{\nabla_{\mathbf{x}} V}(\mathbf{x}) = \sum_{\ell=1}^{M} \pi_{\ell}\,(\mathbf{x}) \nabla_{\mathbf{x}} V(\mathbf{x} + \boldsymbol{\nu}_{\ell}). \tag{S24}$$

Taking the expectation of the PST sample gradient over the hard reaction gives

$$\mathbf{g}_{\text{PST}} = \sum_{j=1}^{M} \left[\boldsymbol{\nu}_j^{\mathsf{T}} \overline{\nabla_{\mathbf{x}} V}(\mathbf{x})\right] \nabla_{\boldsymbol{\theta}} \pi_j. \tag{S25}$$

The exact and PST contributions therefore differ by

$$\mathbf{g}_{\text{exact}} - \mathbf{g}_{\text{PST}} = \sum_{j=1}^{M} \left[\Delta_{\boldsymbol{\nu}_j} V(\mathbf{x}) - \boldsymbol{\nu}_j^{\mathsf{T}} \overline{\nabla_{\mathbf{x}} V}(\mathbf{x})\right] \nabla_{\boldsymbol{\theta}} \pi_j. \tag{S26}$$

Equation (S26) gives the central local reaction-selection discrepancy, which can arise at a single categorical node and is propagated along a trajectory. The exact discrete gradient asks how the downstream value changes when the state is moved by the full stoichiometric jump. PST instead propagates an infinitesimal state derivative through the exact sampled post-reaction state and averages the resulting sample gradient.

## S3. Curvature decomposition and sufficient exactness conditions

Add and subtract $\boldsymbol{\nu}_j^{\mathrm{T}}\nabla_{\mathbf{x}}V(\mathbf{x})$ inside the bracket of Eq. (S26):

$$\Delta_{\boldsymbol{\nu}_j}V(\mathbf{x}) - \boldsymbol{\nu}_j^{\mathrm{T}}\overline{\nabla_{\mathbf{x}}V}(\mathbf{x}) = R_j^{(1)} + R_j^{(2)}, \tag{S27}$$

where

$$R_j^{(1)} = V(\mathbf{x}+\boldsymbol{\nu}_j) - V(\mathbf{x}) - \boldsymbol{\nu}_j^{\mathrm{T}}\nabla_{\mathbf{x}}V(\mathbf{x}), \tag{S28}$$

and

$$R_j^{(2)} = \boldsymbol{\nu}_j^{\mathrm{T}}\left[\nabla_{\mathbf{x}}V(\mathbf{x}) - \overline{\nabla_{\mathbf{x}}V}(\mathbf{x})\right]. \tag{S29}$$

The first term is the finite-difference remainder. If $V$ is twice continuously differentiable along the segment from $\mathbf{x}$ to $\mathbf{x}+\boldsymbol{\nu}_j$, Taylor's theorem in integral form gives

$$R_j^{(1)} = \int_0^1 (1-t)\,\boldsymbol{\nu}_j^{\mathrm{T}}\nabla_{\mathbf{x}}^2V(\mathbf{x}+t\boldsymbol{\nu}_j)\boldsymbol{\nu}_j\,dt. \tag{S30}$$

The second term is an evaluation-point remainder. Using Eq. (S24),

$$R_j^{(2)} = -\boldsymbol{\nu}_j^{\mathrm{T}}\sum_{\ell=1}^{M}\pi_\ell\left[\nabla_{\mathbf{x}}V(\mathbf{x}+\boldsymbol{\nu}_\ell) - \nabla_{\mathbf{x}}V(\mathbf{x})\right], \tag{S31}$$

and each gradient difference can likewise be written as

$$\nabla_{\mathbf{x}}V(\mathbf{x}+\boldsymbol{\nu}_\ell) - \nabla_{\mathbf{x}}V(\mathbf{x}) = \int_0^1 \nabla_{\mathbf{x}}^2\,V(\mathbf{x}+t\boldsymbol{\nu}_\ell)\boldsymbol{\nu}_\ell\,dt. \tag{S32}$$

Suppose the operator norm of the Hessian is bounded by $H$ on all of these connecting segments. Then

$$|R_j^{(1)}| \le \frac{H}{2}\,\|\,\boldsymbol{\nu}_j\,\|^2, \tag{S33}$$

and

$$|R_j^{(2)}| \leq H \parallel \boldsymbol{\nu}_j \parallel \sum_{\ell=1}^{M} \pi_\ell \parallel \boldsymbol{\nu}_\ell \parallel. \quad \text{(S34)}$$

Consequently,

$$\|\mathbf{g}_{\text{exact}} - \mathbf{g}_{\text{PST}}\| \leq \sum_{j=1}^{M} \|\nabla_{\boldsymbol{\theta}} \pi_j\| \left[ \frac{H}{2} \parallel \boldsymbol{\nu}_j \parallel^2 + H \parallel \boldsymbol{\nu}_j \parallel \sum_{\ell=1}^{M} \pi_\ell \parallel \boldsymbol{\nu}_\ell \parallel \right]. \quad \text{(S35)}$$

This bounds the reaction-selection contribution at the current categorical node. The quantity $H$ depends on the downstream objective, the remaining horizon, and the differentiable extension used by the backward computation.

A necessary and sufficient condition for equality of the two local vector contributions is simply

$$\sum_{j=1}^{M} \left[ \Delta_{\boldsymbol{\nu}_j} V(\mathbf{x}) - \boldsymbol{\nu}_j^{\mathsf{T}} \overline{\nabla_{\mathbf{x}} V}(\mathbf{x}) \right] \nabla_{\boldsymbol{\theta}} \pi_j = \mathbf{0}. \quad \text{(S36)}$$

Because different channel contributions can cancel, channelwise equality is stronger than necessary. A convenient sufficient condition is

$$\Delta_{\boldsymbol{\nu}_j} V(\mathbf{x}) = \boldsymbol{\nu}_j^{\mathsf{T}} \overline{\nabla_{\mathbf{x}} V}(\mathbf{x}) \quad \text{for every channel with } \nabla_{\boldsymbol{\theta}} \pi_j \neq \mathbf{0}. \quad \text{(S37)}$$

An especially transparent structural condition is that $V$ be affine on a neighborhood containing $\mathbf{x}$ and all reachable post-reaction states:

$$V(\mathbf{z}) = \mathbf{c}^{\mathsf{T}} \mathbf{z} + d. \quad \text{(S38)}$$

Then $\nabla_{\mathbf{x}} V(\mathbf{z}) = \mathbf{c}$ is constant,

$$V(\mathbf{x} + \boldsymbol{\nu}_j) - V(\mathbf{x}) = \boldsymbol{\nu}_j^{\mathsf{T}} \mathbf{c} = \boldsymbol{\nu}_j^{\mathsf{T}} \overline{\nabla_{\mathbf{x}} V}(\mathbf{x}), \quad \text{(S39)}$$

so both $R_j^{(1)}$ and $R_j^{(2)}$ vanish. This recovers the exact one-step result for affine post-reaction observables.

For a longer trajectory, the same local replacement occurs at each categorical node with a different continuation value. The corresponding errors need not add with a fixed sign: they can be transported, amplified, damped, or canceled by later dynamics and by the objective.

## S4. Worked two-step example

Let $\theta > 0, \beta > 0$, and let the fixed initial state satisfy $X_0 = x \in \mathbb{Z}_{\geq 2}$. The restriction $x \geq 2$ makes the first reaction nondegenerate and ensures that the continuous extension used below has a strictly positive denominator throughout the segment $[x, x+1]$. Consider two competing channels. Channel 1 increments the scalar state by one and has propensity

$$a_1(X;\theta) = \theta X(X-1), \tag{S40}$$

whereas channel 2 leaves the state unchanged and has constant propensity

$$a_2(X) = \beta. \tag{S41}$$

The probability of selecting the incrementing channel is

$$p(X;\theta) = \frac{\theta X(X-1)}{\theta X(X-1)+\beta}. \tag{S42}$$

Starting from the fixed state $X_0 = x$, let

$$K_1 \sim \text{Bernoulli}(p(x;\theta)), \qquad X_1 = x + K_1, \tag{S43}$$

followed by

$$K_2 \sim \text{Bernoulli}(p(X_1;\theta)), \qquad X_2 = X_1 + K_2. \tag{S44}$$

Introduce the shorthand

$$p_0 = p(x;\theta), \qquad p_1 = p(x+1;\theta). \tag{S45}$$

The exact mean final state is

$$m(\theta) = \mathbb{E}[X_2] = x + p_0 + (1-p_0)p_0 + p_0 p_1. \tag{S46}$$

The derivative can be separated into three contributions:

$$\frac{dm}{d\theta} = p_\theta(x) + p_\theta(x)[p_1 - p_0] + \mathbb{E}[p_\theta(X_1)]. \tag{S47}$$

The first term is the exact one-step sensitivity of $\mathbb{E}[K_1]$. The last term is the direct parameter derivative of the second-step probability at the hard state $X_1$. The middle term is the influence of the first categorical choice on the second-step probability. It contains the finite difference

$$p(x+1;\theta) - p(x;\theta). \tag{S48}$$

Under PST, the hard forward value of $X_1$ is retained, but its backward derivative is

$$\frac{dX_{1,\text{PST}}}{d\theta} = p_\theta(x). \tag{S49}$$

The expected PST derivative is therefore

$$\frac{\widetilde{dm}}{d\theta} = p_\theta(x) + p_\theta(x)\mathbb{E}[p_x(X_1;\theta)] + \mathbb{E}[p_\theta(X_1;\theta)]. \tag{S50}$$

Subtracting Eq. (S50) from Eq. (S47) gives the complete discrepancy for this two-step mean:

$$\frac{dm}{d\theta} - \frac{\widetilde{dm}}{d\theta} = p_\theta(x)\{p(x+1;\theta) - p(x;\theta) - \mathbb{E}[p_x(X_1;\theta)]\}. \tag{S51}$$

Because $X_1 = x$ with probability $1 - p_0$ and $X_1 = x + 1$ with probability $p_0$,

$$\mathbb{E}[p_x(X_1;\theta)] = (1 - p_0)p_x(x;\theta) + p_0 p_x(x+1;\theta). \tag{S52}$$

The exact finite difference has the integral representation

$$p(x+1;\theta) - p(x;\theta) = \int_0^1 p_x\,(x+t;\theta)\,dt. \tag{S53}$$

Equations (S52)–(S53) can be viewed as a quadrature mismatch: the exact finite difference averages $p_x$ uniformly along the unit stoichiometric segment, whereas PST uses a probability-weighted average at the two discrete post-reaction states. Their difference is a concrete instance of the two remainders in Eqs. (S28)–(S29); this interpretation is specific to the two-channel example and is not a general balanced-propensity criterion.

For the continuous extension used here,

$$p_\theta(X;\theta) = \frac{\beta X(X-1)}{[\theta X(X-1) + \beta]^2}, \tag{S54}$$

and

$$p_x(X;\theta) = \frac{\theta\beta(2X-1)}{[\theta X(X-1) + \beta]^2}. \tag{S55}$$

The equality in Eq. (S51) can hold at particular states or parameter values even though $p$ is nonlinear, and weighted cancellations can occur in larger systems.

## S5. Waiting times and time-dependent observables

At a fixed state, the SSA waiting time is exactly reparameterizable:

$$\Delta t = -\frac{\log U}{a_0(\mathbf{x};\boldsymbol{\theta})}, \qquad U \sim \mathrm{Uniform}(0,1). \tag{S56}$$

Holding $\mathbf{x}$ fixed,

$$\nabla_{\boldsymbol{\theta}}\Delta t = -\frac{\Delta t}{a_0(\mathbf{x};\boldsymbol{\theta})}\nabla_{\boldsymbol{\theta}}a_0(\mathbf{x};\boldsymbol{\theta}), \tag{S57}$$

which is the exact pathwise derivative of the conditional exponential variate. PST is not needed for this direct derivative.

A later waiting time, however, is evaluated at a state produced by previous categorical choices. For a fixed subsequent uniform variate $U'$ define

$$g_{U'}(\mathbf{z}) = -\frac{\log U'}{a_0(\mathbf{z};\boldsymbol{\theta})}. \tag{S58}$$

The exact contribution of the current reaction probabilities to the expected later waiting time is

$$g_{\text{exact}}^{(U')}(\boldsymbol{z}) = \sum_{j=1}^{M}\left(\nabla_{\boldsymbol{\theta}}\pi_j\right)\left[g_{U'}(\mathbf{x}+\boldsymbol{\nu}_j) - g_{U'}(\mathbf{x})\right]. \tag{S59}$$

The expected PST contribution is

$$g_{\text{PST}}^{(U')}(\boldsymbol{z}) = \sum_{j=1}^{M}\left(\nabla_{\boldsymbol{\theta}}\pi_j\right)\boldsymbol{\nu}_j^{\mathsf{T}}\sum_{\ell=1}^{M}\pi_\ell\,\nabla g_{U'}(\mathbf{x}+\boldsymbol{\nu}_\ell). \tag{S60}$$

Equations (S59)–(S60) have exactly the same finite-difference versus directional-derivative structure as Eqs. (S20) and (S25). Averaging over $U'$ replaces $g_{U'}(\mathbf{z})$ by the conditional mean waiting time $1/a_0(\mathbf{z})$ and leaves the same conclusion unchanged.

## S6. Exact mean matches remain stationary under the PST update field

Let $\mathbf{m}(\boldsymbol{\theta}) \in \mathbb{R}^q$ be a vector of population observables generated by the exact forward simulator, let $\boldsymbol{\mu}$ be the target, and consider

$$\mathcal{L}(\boldsymbol{\theta}) = \frac{1}{2}[\mathbf{m}(\boldsymbol{\theta}) - \boldsymbol{\mu}]^{\mathsf{T}}\mathbf{W}[\mathbf{m}(\boldsymbol{\theta}) - \boldsymbol{\mu}], \tag{S61}$$

where $\mathbf{W}$ is symmetric positive semidefinite. If

$$\mathbf{J}_m = \frac{\partial \mathbf{m}}{\partial \boldsymbol{\theta}} \tag{S62}$$

is the exact sensitivity matrix, then

$$\nabla_{\boldsymbol{\theta}}\mathcal{L} = \mathbf{J}_m^{\mathsf{T}}\mathbf{W}(\mathbf{m} - \boldsymbol{\mu}). \tag{S63}$$

Let $\tilde{\mathbf{J}}_m$ denote the sensitivity supplied by PST while the forward computation of $\mathbf{m}$ remains exact. The update field used by PST is

$$\mathbf{g}_{\mathrm{PST}} = \tilde{\mathbf{J}}_m^{\mathrm{T}} \mathbf{W}(\mathbf{m} - \boldsymbol{\mu}). \tag{S64}$$

Every exact mean match, $\mathbf{m}(\boldsymbol{\theta}) = \boldsymbol{\mu}$, is therefore stationary under both Eqs. (S63) and (S64). Additional stationary points can be introduced by PST when

$$\tilde{\mathbf{J}}_m^{\mathrm{T}} \mathbf{W}(\mathbf{m} - \boldsymbol{\mu}) = \mathbf{0} \tag{S65}$$

despite a nonzero exact gradient. A useful local diagnostic is the alignment

$$\langle \nabla_{\boldsymbol{\theta}} \mathcal{L}, \mathbf{g}_{\mathrm{PST}} \rangle. \tag{S66}$$

A positive value means the negative PST direction is a descent direction for the exact population loss at that point.

For a finite trajectory ensemble, let $\hat{\mathbf{m}}$ be the sampled mean, let $\hat{\tilde{\mathbf{J}}}_m$ be its PST sensitivity, and construct the sampled loss with the same fixed matrix $\mathbf{W}$. The stochastic update retains the exact algebraic residual factorization,

$$\hat{\mathbf{g}}_{\mathrm{PST}} = \hat{\tilde{\mathbf{J}}}_m^{\mathrm{T}} \mathbf{W}(\hat{\mathbf{m}} - \boldsymbol{\mu}). \tag{S67}$$

Both the residual and the sensitivity are Monte Carlo estimates computed from the sampled ensemble.

## S7. Backward jump-operator, small-jump, and system-size interpretation

The finite difference in Eq. (S20) is the same jump increment that appears in the backward generator of a Markov jump process. For a state function $V$, the full continuous-time generator is

$$(\mathcal{L}_{\boldsymbol{\theta}} V)(\mathbf{x}) = \sum_{j=1}^{M} a_j\,(\mathbf{x}; \boldsymbol{\theta}) \big[ V(\mathbf{x} + \boldsymbol{\nu}_j) - V(\mathbf{x}) \big]. \tag{S68}$$

Equation (S20) has the same step-operator increment but a different prefactor, $\nabla_{\boldsymbol{\theta}} \pi_j$, because the derivation isolates the sensitivity of reaction identity conditional on an event. The total-rate contribution, the current waiting-time derivative, and direct parameter dependence of $V$ are treated separately. PST therefore does not truncate the complete generator and does not replace the exact forward process. The shared finite difference nevertheless connects the local reaction-selection analysis to classical Kramers–Moyal small-jump expansions (7, 8).

To make the connection explicit, introduce a formal jump-size parameter $\varepsilon$ and write $\boldsymbol{\nu}_j = \varepsilon \mathbf{s}_j$. Define

$$\bar{\mathbf{s}} = \sum_{\ell=1}^{M} \pi_\ell\, \mathbf{s}_\ell, \qquad \mathbf{H} = \nabla_{\mathbf{x}}^2 V(\mathbf{x}). \tag{S69}$$

Assume that $V$ is three times continuously differentiable on the relevant neighborhood. Taylor expansion gives

$$V(\mathbf{x}+\varepsilon\mathbf{s}_j)-V(\mathbf{x})=\varepsilon\mathbf{s}_j^{\mathsf{T}}\nabla_{\mathbf{x}}V(\mathbf{x})+\frac{\varepsilon^2}{2}\mathbf{s}_j^{\mathsf{T}}\mathbf{H}\mathbf{s}_j+O(\varepsilon^3). \tag{S70}$$

The probability-weighted post-reaction gradient satisfies

$$\overline{\nabla_{\mathbf{x}}V}(\mathbf{x})=\nabla_{\mathbf{x}}V(\mathbf{x})+\varepsilon\mathbf{H}\bar{\mathbf{s}}+O(\varepsilon^2), \tag{S71}$$

and therefore

$$\varepsilon\mathbf{s}_j^{\mathsf{T}}\overline{\nabla_{\mathbf{x}}V}(\mathbf{x})=\varepsilon\mathbf{s}_j^{\mathsf{T}}\nabla_{\mathbf{x}}V(\mathbf{x})+\varepsilon^2\mathbf{s}_j^{\mathsf{T}}\mathbf{H}\bar{\mathbf{s}}+O(\varepsilon^3). \tag{S72}$$

The channelwise discrepancy is thus

$$\begin{aligned}&V(\mathbf{x}+\varepsilon\mathbf{s}_j)-V(\mathbf{x})-\varepsilon\mathbf{s}_j^{\mathsf{T}}\overline{\nabla_{\mathbf{x}}V}(\mathbf{x})\\&\quad=\varepsilon^2\left[\frac{1}{2}\mathbf{s}_j^{\mathsf{T}}\mathbf{H}\mathbf{s}_j-\mathbf{s}_j^{\mathsf{T}}\mathbf{H}\bar{\mathbf{s}}\right]+O(\varepsilon^3).\end{aligned} \tag{S73}$$

PST is therefore locally first-order accurate in stoichiometric jump size, while its leading local discrepancy is quadratic. The first second-order term is the self-contraction familiar from the second jump moment; the second arises because PST averages gradients evaluated at the exact post-reaction states. Both vanish when the continuation value is affine.

The aggregate second-order term has a compact covariance form. Let

$$\mathrm{Cov}_{\boldsymbol{\pi}}(\mathbf{s}_J)=\sum_{j=1}^{M}\pi_j\,\mathbf{s}_j\mathbf{s}_j^{\mathsf{T}}-\bar{\mathbf{s}}\,\bar{\mathbf{s}}^{\mathsf{T}}, \tag{S74}$$

and let $\theta_r$ be one component of $\boldsymbol{\theta}$. For parameter-independent stoichiometry, with direct parameter dependence of $V$ kept in Eq. (S17), Eq. (S73) gives

$$g_{\mathrm{exact},r}-g_{\mathrm{PST},r}=\frac{\varepsilon^2}{2}\mathbf{H}\colon\frac{\partial}{\partial\theta_r}\mathrm{Cov}_{\boldsymbol{\pi}}(\mathbf{s}_J)+O(\varepsilon^3), \tag{S75}$$

where $\mathbf{A}\colon\mathbf{B}=\mathrm{tr}(\mathbf{A}^{\mathsf{T}}\mathbf{B})$. This expression separates an objective-side factor, the downstream Hessian, from a network-side factor, the parameter sensitivity of the conditional jump covariance. It is at the same jump-moment order as diffusion corrections in a Kramers–Moyal expansion, but it is not the chemical diffusion tensor: the latter contains the event-rate-weighted raw second moment, whereas Eq. (S75) concerns the conditional reaction identity and its parameter derivative.

A system-size interpretation follows directly from the nonasymptotic bound in Eq. (S35). Under standard density-dependent scaling (9, 10), write

$$\mathbf{x}=\Omega\mathbf{z},\qquad a_j^{\Omega}(\Omega\mathbf{z};\boldsymbol{\theta})=\Omega\alpha_j(\mathbf{z};\boldsymbol{\theta}),\qquad V^{\Omega}(\mathbf{x})=\Omega v\left(\frac{\mathbf{x}}{\Omega}\right), \tag{S76}$$

with a fixed finite set of $O(1)$ stoichiometric vectors. Then

$$\nabla_{\mathbf{x}} V^{\Omega} = \nabla_{\mathbf{z}} v, \qquad \nabla_{\mathbf{x}}^2 V^{\Omega} = \Omega^{-1} \nabla_{\mathbf{z}}^2 v, \qquad \pi_j^{\Omega} = \frac{\alpha_j}{\sum_{\ell} \alpha_{\ell}}. \tag{S77}$$

If the stoichiometric norms, $\nabla_{\boldsymbol{\theta}} \pi_j^{\Omega}$, and $\nabla_{\mathbf{z}}^2 v$ are uniformly bounded, Eq. (S35) yields

$$\|\mathbf{g}_{\text{exact}}^{\Omega} - \mathbf{g}_{\text{PST}}^{\Omega}\| = O(\Omega^{-1}) \tag{S78}$$

at one reaction-selection node. When the leading exact local contribution has a nonzero $O(1)$ limit, the corresponding local relative discrepancy is also $O(\Omega^{-1})$. Thus, PST reproduces the leading macroscopic reaction-selection sensitivity and becomes locally asymptotically exact in the small-concentration-jump limit. The scaling in Eq. (S78) is the same large-system scaling that underlies the van Kampen system-size expansion and the density-dependent limit. PST, however, is not derived by expanding the probability law around a deterministic trajectory; the forward jump process remains exact at every Ω. Nevertheless, its curvature-controlled local discrepancy is suppressed by exactly the mechanism on which those classical limits rest. Small jumps are sufficient but not necessary for accuracy: the affine-continuation condition in Eq. (S38) gives exact equality at any jump size.

## S8. Relation to score-function, marginal straight-through, and Gumbel-Softmax estimators

For channels with positive probability, the exact reaction-selection contribution in Eq. (S20) has the score-function form

$$\mathbf{g}_{\text{exact}} = \mathbb{E}_{J\sim\boldsymbol{\pi}}\left[\nabla_{\boldsymbol{\theta}} \log \pi_J \left(V(\mathbf{x} + \boldsymbol{\nu}_J) - V(\mathbf{x})\right)\right]. \tag{S79}$$

Thus $V(\mathbf{x})$ is a valid baseline and the finite post-reaction value difference is an advantage-like quantity. PST replaces this exact finite difference with a straight-through directional-derivative surrogate. This identity connects the local comparison to likelihood-ratio and policy-gradient methods (11).

Computational cost per gradient and the trajectory budget needed for useful updates are distinct aspects of scalability. For a scalar trajectory loss, a score-function estimator uses the trajectory score vector multiplied by a scalar loss or baseline-adjusted return. The score vector can be computed by reverse-mode differentiation of the trajectory log likelihood with the sampled path held fixed; separate simulations for each parameter are not intrinsic to this construction. Its sampling cost nevertheless depends on controlling gradient noise. Standard parameterwise coupled finite-difference and Poisson-path constructions, discussed in Ref. (1), instead incur additional parameter-specific simulations or auxiliary sensitivity calculations. In the reverse-mode implementations used for large models, GS-ST and PST accumulate surrogate gradients through vector–Jacobian products on the trajectory graph, avoiding parameterwise simulation replicas. Costs still depend on event count, propensity evaluation, graph storage or recomputation, and ensemble size. Unbiasedness alone specifies neither computational nor sampling cost. The classifier demonstrates PST training at 203,796 parameters within the established GS-ST framework. The present experiments do not benchmark score-function or coupling-based estimators on that classifier and do not establish a universal inability of those approaches to scale.

The SSA-specific contribution of PST is the combination of exact categorical marginals with the affine stoichiometric update: this makes the derivative exact for the one-step conditional mean and affine post-reaction observables and yields the explicit multistep discrepancy derived above. For categorical sampling ($k = 1$), those marginals reduce to the normalized category probabilities. The marginal construction is therefore shared, whereas the one-step exactness result and the trajectory-level analysis arise from the SSA update structure.

Gumbel-Softmax straight-through estimation uses an exact hard categorical reaction in the forward pass but differentiates a relaxed sample

$$\tilde{\mathbf{y}}_{T,G} = \mathrm{softmax}\left(\frac{\log \boldsymbol{a} + \mathbf{G}}{T}\right), \tag{S80}$$

where $\mathbf{G}$ contains Gumbel variates and $T$ is a temperature (1, 12, 13). PST shares the hard forward reaction selection but replaces the temperature-dependent relaxed backward sample with

$$\boldsymbol{\pi} = \frac{\mathbf{a}}{\sum_j a_j}. \tag{S81}$$

For positive propensities, $\mathrm{softmax}(\log \boldsymbol{a}) = \boldsymbol{\pi}$. This identity permits existing exact-forward GS-ST code to retain the hard categorical sample while replacing the backward relaxation by normalized propensities. PST is not equivalent to GS-ST at $T = 1$: GS-ST retains the random Gumbel perturbation and its temperature-dependent derivative. A shared softmax code path therefore supports the cost-matched construction used in the main-text oscillator comparison: removing the backward Gumbel perturbation and using unit scaling returns PST while retaining the same hard forward sampling law. Optimization-step timing benchmarks used a common hard Gumbel-Max sampler for PST and GS-ST on an NVIDIA RTX 6000 Ada Generation GPU with TensorFlow 2.20.0. The workloads comprised 204,800 trajectories of 150 events for the oscillator and 262,144 trajectories of 20 event slots for the ion-channel model. After three warm-up calls per method, 18 paired repeats used balanced execution order. Timings included forward simulation, readout and loss, reverse-mode differentiation, and the optimizer update; tracing, resets, reference generation, and oscillator batch selection were excluded. The PST/GS-ST ratios of median optimization-step times were 1.002 (paired-bootstrap 95% interval 0.943–1.038) for the oscillator and 0.985 (0.962–0.999) for the ion-channel model, using 20,000 paired bootstrap resamples.

The mean of the Gumbel-Softmax relaxation provides a complementary connection. The Gumbel-perturbed logits have a unique maximizer almost surely, so

$$\tilde{\mathbf{y}}_{T,G} \underset{T\to 0^+}{\overset{\text{a.s.}}{\to}} \mathbf{e}_J, \qquad J = \arg\max_j \left(\log a_j + G_j\right). \tag{S82}$$

Because the components of $\tilde{\mathbf{y}}_{T,G}$ are bounded and the Gumbel-Max identity gives $J \sim \mathrm{Categorical}(\boldsymbol{\pi})$, bounded convergence gives

$$\lim_{T\to 0^+} \mathbb{E}_{\mathbf{G}}\left[\tilde{\mathbf{y}}_{T,G}\right] = \boldsymbol{\pi}. \tag{S83}$$

The normalized propensities are therefore both the exact categorical mean and the zero-temperature limit of the mean of the Gumbel-Softmax relaxation. PST is derived directly from the categorical-mean identity; the limit is included only to clarify the relation to GS-ST.

## S9. Summary

For one categorical SSA reaction followed by a downstream continuation value, the exact reaction-selection gradient is weighted by the same finite-difference increment that appears in the backward jump generator. At the affine one-step state update, the PST reaction-selection derivative is independent of the realized reaction index and therefore has zero reaction-selection variance conditional on the current state. PST preserves the exact hard forward state but replaces this finite jump action by directional derivatives of the continuous backward extension, averaged over the exact post-reaction distribution. For nonlinear continuation values, a discrepancy can already arise after one reaction. The leading local small-jump discrepancy is quadratic under the smoothness and boundedness conditions developed above and separates downstream curvature from the parameter sensitivity of the conditional jump covariance. Two distinct regimes therefore suppress the local error: exact cancellation for affine continuation values, irrespective of jump size, and asymptotic reduction when individual concentration jumps become small under density-dependent scaling while the relevant curvature and probability sensitivities remain bounded. The same finite-difference-to-directional-derivative mechanism applies recursively over multiple reactions and to later waiting times and readout operations. The two-step example in Section S4 provides an explicit closed-form instance of the replacement without invoking a relaxed-forward construction.

# Supplementary Analyses

## Objective dependence in the repressilator benchmark

*Controlled loss comparison*

The main-text repressilator analysis uses relative mean-squared error with a fixed target scale. To assess whether recovery depends on the objective, we repeated the complete 50-task calculation with the log-of-ensemble-mean loss used by Burger et al. (6). The two saved protocol records differ only in `loss_mode` (apart from the identifying code-version string): the exact simulator, task array, initial conditions, observation times, 50,000 reference trajectories, 50,000 training trajectories, 50,000 validation trajectories, PST and GS-ST definitions, temporal straight-through readout, RMSprop settings, global gradient clipping, learning-rate schedule, SNR stopping, and refinement stage are unchanged.

For one species–time component, the log-of-ensemble-mean objective of Ref. (6) is

$$\ell_{\log} = [\log \hat{\mu} - \log \mu^*]^2, \tag{S84}$$

where $\hat{\mu}$ is the trainable ensemble mean and $\mu^*$ is the fixed target mean. Exactly zero means are replaced numerically by $10^{-20}$, matching the implementation used in the supplementary run. For positive $\hat{\mu}$, its derivative is

$$\frac{\partial \ell_{\log}}{\partial \hat{\mu}} = \frac{2\log(\hat{\mu}/\mu^*)}{\hat{\mu}}. \tag{S85}$$

The relative objective used in the main analysis is

$$\ell_{\text{rel}} = \left(\frac{\hat{\mu} - \mu^*}{\mu^*}\right)^2, \tag{S86}$$

with derivative

$$\frac{\partial \ell_{\mathrm{rel}}}{\partial \hat{\mu}} = \frac{2(\hat{\mu} - \mu^*)}{(\mu^*)^2}. \tag{S87}$$

The distinction relevant to low-copy observations is that Eq. (S85) divides by the fluctuating trainable mean, whereas Eq. (S87) uses the fixed target in its denominator. The resulting task-level recovery summaries are reported in Table S1.

**Table S1. Parameter recovery under the two repressilator objectives.** MAPE is the mean absolute percentage error over the inferred pair $(k_p, K_d)$, evaluated separately for each task. The success count requires both inferred parameters to lie within $20\%$ of their reference values. Each objective uses one seeded optimization realization per method and task.

| Objective | Method | Median task MAPE | Mean task MAPE | Maximum task MAPE | Both parameters within $20\%$ |
|---|---|---|---|---|---|
| Relative mean-squared error | PST | $0.135\%$ | $0.381\%$ | $5.677\%$ | 50/50 |
| Relative mean-squared error | GS-ST | $0.137\%$ | $0.378\%$ | $5.771\%$ | 50/50 |
| Log of ensemble mean | PST | $0.165\%$ | $9.375\%$ | $248.458\%$ | 48/50 |
| Log of ensemble mean | GS-ST | $0.110\%$ | $4.307\%$ | $197.067\%$ | 49/50 |

The median task errors remain below 0.2% for both estimators under either objective. The principal effect is instead concentrated in the tail of the task distribution. Under the log objective, PST fails the $20\%$ criterion on tasks 34 and 42, and GS-ST fails on task 42. Task 42 is therefore a common failure of both straight-through estimators. Its reference parameters are

$$(k_p^*, K_d^*) = (30.269,\ 0.7537). \tag{S88}$$

The log-objective estimates are $(10.684, 4.011)$ for PST and $(11.219, 3.250)$ for GS-ST, placing both methods in the same incorrect region. Under relative mean-squared error, the corresponding estimates are $(30.236, 0.7508)$ and $(30.240, 0.7566)$, respectively. Task 34 is an estimator-specific outlier under the log objective: PST returns $(9.396, 3.143)$ for the reference pair $(22.954, 0.7054)$, whereas GS-ST returns $(22.955, 0.7019)$.

These data show that PST and GS-ST achieve the same practical parameter-recovery accuracy for the large majority of tasks under either loss. Under this fixed adapted protocol, rare catastrophic recovery failures were observed with the log-of-ensemble-mean objective but not with relative mean-squared error. The result is not a direct reproduction of the original Burger optimization experiment, which used a different optimizer and trajectory budget. The controlled comparison establishes that changing the objective eliminated the observed recovery failures for both reaction-selection estimators in this protocol. The result makes objective choice a useful design diagnostic when trainable ensemble means can approach zero. The representative fit, paired task errors, and parameter-recovery patterns under the alternative objective are shown in Figure S1.

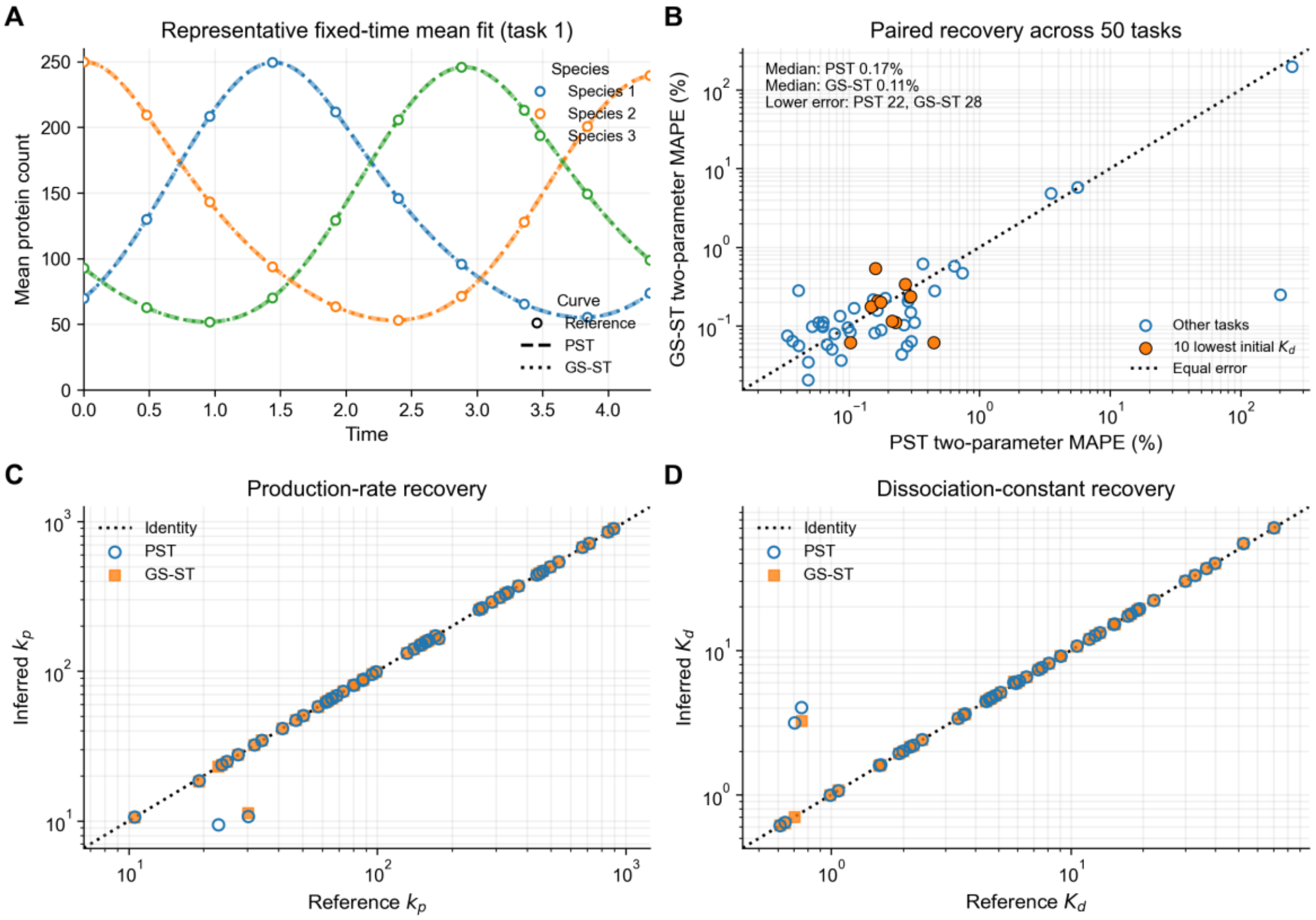

**Figure S1. Repressilator inference with the log-of-ensemble-mean objective of Ref.** (6)**.** The model, 50 reference–initialization pairs, simulator, trajectory counts, optimizer, stopping rule, and refinement stage are the same as in the relative-mean-squared-error analysis of main-text Figure 6; only the loss is changed. **A**, Representative fixed-time ensemble-mean fit for task 1. Open markers show exact-SSA reference means; dashed and dotted curves show independent exact-SSA validation means at the PST- and GS-ST-inferred parameter pairs. **B**, Paired task-wise two-parameter MAPE for PST and GS-ST. The diagonal denotes equal error, and filled symbols identify the ten tasks with the lowest initial $K_d$. **C**, Inferred production rate $k_p$ versus its reference value. **D**, Inferred dissociation constant $K_d$ versus its reference value. The log objective retains median MAPEs of 0.165% for PST and 0.110% for GS-ST but produces catastrophic outliers: both parameters are within 20% in 48/50 PST tasks and 49/50 GS-ST tasks. Relative mean-squared error restores this criterion to 50/50 tasks for both methods (main-text Figure 6 and Table S1).

## Training the established stochastic classifier with PST

### *PST implementation of the stochastic MNIST classifier*

We tested PST at deep-learning parameter scale using the stochastic gene-regulatory classifier previously introduced and optimized in Ref. (1). The network contains 203,796 trainable parameters and maps 784 input transcription-factor concentrations through 256 hidden genes to 10 output genes. The architecture, MNIST data split, stochastic reaction-network dynamics, categorical cross-entropy objective, RMSprop optimizer, stochastic weight averaging (SWA), and post-training Monte Carlo and temporal-averaging evaluation follow the earlier implementation. The reported training realization uses global seed 13 and a different schedule from the earlier GS-ST run, as detailed next.

The original GS-ST calculation used 80 epochs, with soft-forward Gumbel-Softmax sampling during the first 16 epochs, followed by hard-forward straight-through training, and annealed the reaction-selection temperature from $2.0$ to $0.2$ (1). In the PST calculation, training uses 40 epochs and PST is active from epoch 0. Every training forward pass samples an exact hard categorical event, while the backward pass differentiates the normalized propensities.

Training and test losses decrease smoothly, and the recorded accuracy curves improve rapidly through approximately epoch 20 and more gradually thereafter over the 40-epoch history (Figure S2A,B). The final single-pass test accuracy was $97.43\%$. Post-training evaluation used the SWA-averaged weights and combined independent stochastic realizations with averaging over a terminal fraction of each trajectory. Across the complete $6 \times 6$ evaluation grid shown in Figure S2C, accuracy ranged from $97.49\%$ to $98.22\%$. The highest observed value, $98.22\%$, was obtained with four Monte Carlo realizations and averaging over the terminal $64\%$ of the trajectory. These settings and outcomes are summarized in Table S2.

**Table S2. PST training on the established deep-learning-scale architecture.** Summary of the 203,796-parameter stochastic classifier trained with PST from epoch 0 using exact hard categorical reaction selection.

| **Quantity** | **PST calculation** |
|---|---|
| Trainable parameters | 203,796 |
| Training epochs | 40 |
| Global training seed | 13 |
| Reaction-selection rule | PST from epoch 0 |
| Forward simulation during training | Exact hard categorical events |
| Reaction-selection relaxation temperature | None |
| Soft-forward warm-up | None |
| Final single-pass test accuracy | $97.43\%$ |
| Highest observed Monte Carlo and temporally averaged accuracy | $98.22\%$ |
| Setting for highest observed accuracy | 4 realizations; terminal fraction $0.64$ |

The calculations are individual realizations with different schedules, durations, and stochastic backward rules: the earlier GS-ST run used 80 epochs, an initial soft-forward phase, and a temperature schedule. PST reaches an apparent performance plateau within 40 epochs and demonstrates hard-forward optimization at the 203,796-parameter scale without a reaction-selection relaxation temperature or soft-forward initialization.

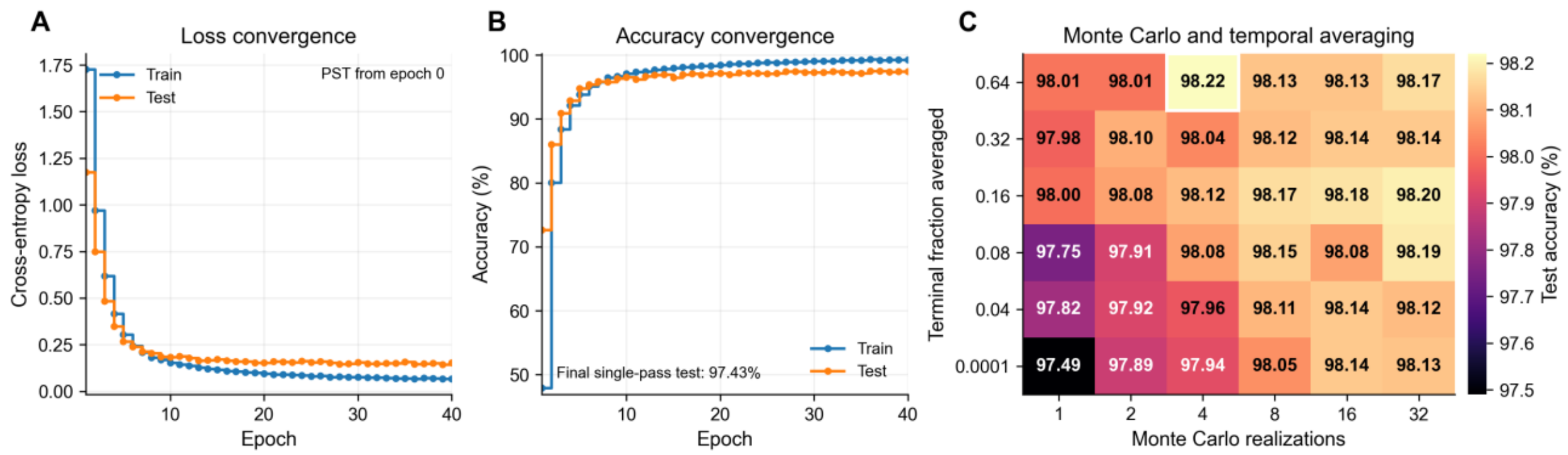

**Figure S2. PST training of the previously established stochastic gene-regulatory classifier.** The network contains 203,796 trainable parameters and was trained for 40 epochs. PST was used from the first epoch, with exact hard categorical reaction selection in every forward pass and no reaction-selection relaxation temperature or soft-forward warm-up. **A,** Training and test cross-entropy losses. **B,** Training and single-pass test accuracy during optimization; the accuracy improves rapidly through approximately epoch 20 and more gradually thereafter, with final single-pass accuracy 97.43%. **C,** Post-training test accuracy using SWA-averaged weights as a function of the number of independent Monte Carlo realizations and the terminal fraction of the stochastic trajectory included in temporal averaging. The highest observed accuracy was 98.22%, obtained with four Monte Carlo realizations and averaging over the terminal 64% of the trajectory. The white outline marks this maximum.

## Sensitivity of the ion-channel loss-stabilization diagnostic

*Terminal-loss bands*

Main-text Figure 4C compares the saved PST and GS-ST training-loss histories on a common scale. For each method, the loss is smoothed with a trailing 10-epoch arithmetic mean. The terminal reference is the arithmetic mean over the final 50 epochs. For a relative tolerance $\delta$, the stabilization epoch is the first completed epoch after which the smoothed loss remains no greater than $(1 + \delta)$ times that terminal reference for the remainder of training.

**Table S3. Sensitivity of the ion-channel stabilization epoch to the terminal-loss band.** Ratios use the GS-ST epoch divided by the PST epoch. All values come from one saved optimization realization per method and are descriptive.

| Terminal-loss band | PST epoch | GS-ST epoch | GS-ST/PST ratio |
|---|---|---|---|
| 20% | 69 | 152 | 2.20 |
| 10% | 72 | 156 | 2.17 |
| 5% | 75 | 160 | 2.13 |
| 2% | 77 | 184 | 2.39 |

The terminal 50-epoch mean losses are $4.3140 \times 10^{-4}$ for PST and $4.3477 \times 10^{-4}$ for GS-ST, a difference of 0.78%. The stabilization ratio remains close to two across a tenfold change in tolerance.

## Exploratory repressilator subset analysis

The small difference in overall median MAPE is not uniform across the 50-task suite. Among the ten tasks with the lowest initial $K_d$ values highlighted in main-text Figure 6B, median MAPE is 0.167% for PST and 0.176% for GS-ST, and each method has lower error in five tasks. Across the remaining forty tasks, the medians are 0.114% for PST and 0.136% for GS-ST.

## Consolidated numerical comparisons

The primary numerical comparisons are consolidated in Table S4.

**Table S4. Primary numerical comparisons.** Oscillator MAPE entries are medians of run-wise MAPEs, with run-level percentile bootstrap intervals. Dimerization errors are mean condition-level MAPEs across eight kinetic conditions. Repressilator values are medians across 50 task-wise two-parameter MAPEs. Ion-channel NRMSE is normalized by the experimental mean-trace range. The oscillator convergence value is a ratio of method-level median threshold epochs. The Figure 5 ion-channel value is the median of 30 within-comparison GS-ST/PST ratios and therefore need not equal the ratio of the two marginal median epochs.

| System and horizon | PST | GS-ST | Interpretation |
|---|---|---|---|
| Reversible dimerization | MAPE 0.060% | MAPE 0.093% | Comparable recovery across the eight conditions |
| Genetic oscillator, 1000 epochs | Run-wise MAPE 1.70% [1.49–1.99%] | 2.58% [1.47–3.78%] | PST reaches useful accuracy earlier |
| Genetic oscillator, 3000 epochs | 1.71% [1.49–1.97%] | 1.51% [1.30–2.34%] | GS-ST has the lower point estimate; intervals overlap |
| Ion channel, Figure 4 | $R^2 = 0.987$, NRMSE 3.49%; 5% loss-stabilization epoch 75 | $R^2 = 0.987$, NRMSE 3.48%; 5% loss-stabilization epoch 160 | Near-identical final fits; the GS-ST/PST stabilization-epoch ratio is 2.13 in one matched-protocol pair |
| Ion channel, Figure 5 reduced ensemble | Median convergence epoch 36.0; median NRMSE 3.46% | Median convergence epoch 90.5; median NRMSE 3.63% | Median paired GS-ST/PST epoch ratio 2.41; PST earlier in 30/30 matched comparisons |
| Repressilator, 50 tasks | Median MAPE 0.13%; both parameters within 20% in 50/50 tasks | Median MAPE 0.14%; both parameters within 20% in 50/50 tasks | Accurate recovery across the fixed task suite; the median difference is descriptive |

# SI References